\documentclass[manuscript]{aastex}
\shorttitle{The Earth-mass Planet around  $\alpha$ Centauri B}
\shortauthors{Hatzes}
\begin{document}

\title{RADIAL VELOCITY DETECTION OF EARTH-MASS PLANETS IN  THE 
PRESENCE OF ACTIVITY NOISE: THE CASE OF $\alpha$ CENTAURI Bb}

\author {Artie P. Hatzes }
\email{artie@tls-tautenburg.de}
\affil{Th\"uringer Landessternwarte, D - 07778 Tautenburg, Germany}

\begin{abstract}
We present an analysis of the publicly available HARPS 
radial velocity (RV) measurements
for $\alpha$ Cen B, a star hosting
an Earth-mass planet candidate in a 3.24\,day orbit.  The goal is to
devise robust ways of extracting low-amplitude RV signals of low mass
planets in the presence of activity noise. Two approaches were used to remove the
stellar activity signal which dominates the RV variations:
1) Fourier component analysis (pre-whitening), and 2) local trend filtering (LTF) of the
activity using short time windows of the data. The Fourier
procedure results in a signal at $P$ = 3.236\,d and $K$ = 0.42 m\,s$^{-1}$ which is
consistent with the presence of an Earth-mass planet, but the false alarm
probability for this signal is rather high at a few percent. The LTF
results in no significant detection of the planet signal, although
it is possible to detect a marginal planet signal with this method 
using a different
choice of time windows and fitting functions. However, even in this case the 
significance of the 3.24-d signal depends on the details of how a time
window containing only 10\% of the data is filtered.
Both methods should have detected the presence of $\alpha$ Cen Bb  
at a higher significance than is actually seen. 
We also investigated the influence of random noise with a standard deviation
comparable to the HARPS data and sampled in the same way.
The distribution of the noise peaks in the 
period range 2.8 -- 3.3\,d have  a maximum of $\approx$ 3.2\,d and
amplitudes approximately one-half of the 
$K$-amplitude for the planet. 
The presence of the activity signal may boost the velocity amplitude of these
signals to values comparable to the planet.  
It may be 
premature to attribute the 3.24\,day RV variations to an Earth-mass planet.
A better understanding of the noise characteristics in the RV data as well  as
more measurements with better sampling will be needed to confirm this exoplanet.
\end{abstract}


\section{Introduction}
\label{intro}

Precise stellar radial velocity (RV) measurements can currently achieve
a precision better than a few m\,s$^{-1}$ and this has enabled astronomers
to detect planets with masses of just a few Earth masses (e.g.
Tuomi et al. 2013; Mayor et al. 2009). The ability of the RV method
to detect the lower  mass planets ($\approx$ 1 $M_{\oplus}$ or smaller)
hinges not on the measurement error provided by instruments, but rather
on the `error' of the intrinsic stellar variability. With a precision of 
below 1 m\,s$^{-1}$ (Pepe et al. 2011)
RV measurements are approaching the stellar noise floor
of many solar-type stars. 

The RV `jitter'
caused by magnetic activity (spots, plage, changes in the convection pattern,
etc.) is a major source of stellar noise that can hinder the detection of
Earth-mass planets.
For example, a spot coverage of only 0.5\%, typical for the sun
at solar maximum, can induce an RV variation of $\approx$ 0.5 m\,s$^{-1}$
(Saar \& Donahue 1997; Hatzes 2002).  This is the velocity 
amplitude of the stellar reflex motion
caused by an Earth-mass planet  around
Sun-like star with an orbital period of a few days. An Earth-mass
planet 1 AU  from the star will cause an RV motion of a mere
 0.09 m\,s$^{-1}$.

In order to detect Earth-mass planets with RV measurements  around solar-type
stars we must devise ways of overcoming the activity noise.
This will be challenging
as these activity RV variations, depending on the spot 
distribution on the stellar surface, will be modulated by the rotation period, 
$P_{rot}$,  of the star as well
as its higher harmonics ($P_{rot}$/2, $P_{rot}$/3, $P_{rot}$/4, etc.).
Spot 
evolution and activity cycles coupled with the sampling window will add
other frequencies to the power spectrum of the RV variations.
These intrinsic variations must be filtered out in a robust way. 

Dumusque et al (2012; hereafter D2012) demonstrated that
it may be possible to `break the activity barrier' and
detect an Earth-mass planet in the presence of 
stellar activity. The authors used RV measurements to find a 
1.13 $\pm$ 0.09 M$_\oplus$ planet with
a 3.236\,day  orbital period around $\alpha$ Cen B.
The discovery of this low mass planet was challenging.
Alpha Centauri B has a modest level of activity that creates substantial
RV `jitter' compared to the planet signal. The  RV measurements
showed a dominant periodic signal at $\approx$ 38 days with an amplitude of $\approx$
1.5 m\,s$^{-1}$ or at least three times the so-called velocity $K$-amplitude  caused by the planet. 
To remove this activity signal D2012 used a harmonic
filtering approach. In this method RV measurements are selected
using a time interval of a few rotational periods 
and fitting the RV activity variations with sine functions using the rotational frequency,
$\nu_{rot}$,
and its harmonics (2$\nu_{rot}$, 3$\nu_{rot}$, etc.). If the peak in the periodogram had
a false alarm probability less than 10 \% and its period was equal to the rotational
period or one of its  harmonics it was removed.
The harmonic method has been used 
to detect the  RV variations of the transiting rocky planet CoRoT-7b
(Queloz et al. 2009; Ferraz-Mello et al. 2011) as well as to reduce the
activity noise of other planet hosting stars (Boisse et al. 2011).

The RV detection of the rocky planet CoRoT-7b also
demonstrated that a planetary signal can be extracted from
RV measurements dominated by variations due to activity.  
The detection of CoRoT-7b was easier than $\alpha$ Cen Bb for
a number of reasons: 1) although the 
$K$-amplitude of the star due to the companion ($\approx$ 5 m\,s$^{-1}$)
was a factor of two smaller than
the activity variations, it was still larger then the measurement error
of $\approx$ 2 m\,s$^{-1}$. 2) The RV data were taken over a  relatively short
time span of a few months. 3) The  orbital period of the planet
was already known from transit light curves (L\'eger et al. 2009). 
4) Finally, the orbital period was
much smaller than the rotational period of the star by a factor of almost 30.

The detection of $\alpha$ Cen Bb posed more challenges compared
to CoRoT-7b:
1) the stellar $K$-amplitude ($\approx$ 0.5 m\,s$^{-1}$)
is comparable, if not smaller than the measurement error and 
smaller than the intrinsic stellar variations.
2) To detect the planet signal D2012
had to combine  data taken over three years,
thus spot evolution and activity cycles
may be more of a problem than it was for CoRoT-7b. 3) The period of the planetary companion was 
not known a priori and
it requires significantly
more data to detect an unknown period in a time series, particularly when the amplitude is comparable
to the measurement error.
4) The planet period was a relatively long 3.236 days which 
is a factor 
of 10 shorter than the rotational period 
of the star. Alpha Cen B thus represents a different and more difficult case than
CoRoT-7b for the detection of planetary signal in the presence of stellar noise.
As noted by Hatzes (2012), the large number  of the RV measurements
for $\alpha$ Cen B as well as their high quality provides an excellent
data set for astronomers to test their analysis tools for extracting planetary
signals in the presence of stellar activity. 

In this paper we analyze the  HARPS RV data for
$\alpha$ Cen B using several approaches
to filter out the activity noise.  There are several goals to this investigation:
1) to test the effectiveness of various approaches to filtering out the
activity. 2) To check the robustness of the planet signal. If it can be detected with different
methods then we can be more confident of its presence. 3) To obtain 
a better determination of the planet mass (i.e. $K$-amplitude). 
CoRoT-7b demonstrated that the various ways of filtering the activity 
resulted in masses
for the transiting rocky planet that initially differed by factors of four 
(see Hatzes et al. 2011). A different analysis may produce a revised mass
 for $\alpha$ Cen Bb.

In most planet detections using RV data researchers rely primarily on the Lomb-Scargle
periodogram and few use the discrete Fourier transform (DFT) which is
primarily employed by the stellar oscillation community. In this work {\it both} tools
will be employed as each has its purpose. We use the DFT to  understand the velocity
amplitude of various signals that are present in the data and to 
detect and subsequently remove the dominant Fourier components
(sine functions). The Scargle (1982) periodogram is used primarily because 
it gives a measure of the statistical significance of a periodic signal.

Since both the DFT and the Scargle periodogram will be employed in this paper it is worthwhile
commenting about the differences between the two.
In DFT  analysis the power is in units of (m\,s$^{-1}$)$^2$. In this paper
the amplitude is shown so that the reader can
readily assess the velocity amplitude of a signal as well as the 
true velocity noise floor surrounding a peak.  
For a DFT the amplitude of a signal remains
constant regardless of the number of data points or significance of a signal.
For a real signal, the acquisition of additional data does 
not change the amplitude significantly, it merely reduces the overall noise floor
in Fourier space. On the other hand, the power in a Scargle periodogram  is a measure
of the statistical significance of a signal.
For a real signal as you acquire more data the significance of the detection
increases and so  does the Scargle power, but in a non-linear way.
As a rule-of-thumb for the periodograms
shown here, Scargle power, $z$, less than 10 is most likely not a significant signal,
10 $<$ $z$ $<$ 14 is a modestly significant signal that is interesting
and merits more investigation (i.e. an improved analysis or more data), whereas $z$ $>$ 20 is most
likely a true signal. 

In the exoplanet community it has become a common practice to show
periodograms using period for the abscissa. In this work a frequency
scale will be used for two reasons. First, frequency is the natural
units when using a DFT or periodogram (in spite of the name).
Second, the use of period distorts the periodogram making it difficult
to judge the comparative width of features as well as the noise floor surrounding a peak.
When appropriate both  period and frequency will  be given.  Frequency will be given in
units of d$^{-1}$ rather than {\rm Hz} since {\it day} is the unit often
used to to express orbital periods for exoplanets.

\section{The Radial Velocity Data}

The high precision RV measurements used for this analysis are
the ones presented by D2012. These were taken with the HARPS
spectrograph at the 3.6m telescope at La Silla Observatory of the European Southern
Observatory. A total of 459 RV measurements were made between 
2008 February and 2011 July.
The RV measurements had a median photon error of 0.4 m\,s$^{-1}$ and
D2012 estimated a systematic error
of 0.7 m\,s$^{-1}$. Adding these in quadrature 
results in  
a total error of about 0.8 m\,s$^{-1}$. We take this as the
``best case'' estimate of the RV error. 
More details of the data analysis and reduction can be found
in D2012.

The RV measurements for $\alpha$  Centauri B  showed a long term trend that is part of the binary
orbital motion with component A. This orbital motion was removed by 
fitting it with a second order polynomial as was done in D2012.
This should be  adequate
since the rotational period of the star and orbital period of the planet
are both considerably less than the binary orbital period of several
decades.
Throughout the paper  the term ``RV data'' will refer to
the RV measurements after removal of only the binary orbital motion.
The term `RV residuals' will refer to RV data that has all variations
presumably due to activity removed, but with any `planetary' signal 
still in the data.

The HARPS data were taken over four epochs spanning more than three
years. Table 1 lists
the Julian day (JD) span of the epoch, the time span in days, the number of measurements,
$N_{obs}$,
made in that epoch, and the standard deviation, 
$\sigma$,  of the RVs after removing the orbital motion. Throughout this paper JD values
will be given as reduced Julian day (RJD = JD $-$ 2400000).
Epoch 1 has a standard deviation
only slightly more than the estimated error indicating a low level of activity for the star,
as noted by D2012. Epoch 3 has the largest standard deviation which implies a more
active phase of the star.

\section{Results}
\subsection{Fourier Component Analysis Via Pre-whitening}
 
Pre-whitening is a commonly 
used tool for finding multi-periodic signals in time
series data. This
method sequentially finds the dominant Fourier components in a time series and
removes them. 
Traditionally it is employed to derive the frequency spectrum
of oscillating stars (e.g. Garc\'ia Hern\'andez et al. 2009), but it also has 
applications as a means of removing the intrinsic stellar variations due to activity. 
The mathematical 
foundation for this is that sines and cosines form a set of basis functions. You 
can represent most functions as a sum of sine waves with different periods and amplitudes.
In fact, the DFT of a time series merely gives you the amplitude as a function of frequency 
of all the sine functions that are present, including those due to noise.
The trick is to use enough Fourier components to represent adequately the function of
interest (activity in this case) without introducing spurious periods, or 
altering the amplitude and frequency of a signal you are trying to detect. 
So long as artifacts introduced by the multi-component sine fit have frequencies and 
amplitudes different
from our signal of interest, pre-whitening can be a useful tool for detecting weak
periodic signals. In this case we have the advantage
in that we have a priori information
about the signal we are trying to detect. 

The pre-whitening process is similar to the harmonic analysis used by
of D2012 in that both fit the activity signal with multi-sine components. There are,
however, two major differences. First,
in pre-whitening the time span is not restricted to a few rotational periods, but in our case
we use a much longer time span.
Second, the frequencies that are removed by pre-whitening are not restricted to just the rotational
frequency and its harmonics. The strongest peak in the DFT is removed regardless
of its frequency. One can consider harmonic analysis as a more restricted version of pre-whitening.

The case of CoRoT-7b demonstrated the effectiveness of Fourier component analysis in
fitting an activity signal. The RV activity jitter for CoRoT-7  was about a factor of 10
higher than for $\alpha$ Cen B.
 The pre-whitening process resulted in a $K$-amplitude
of 5.5 $\pm$ 0.3 m\,s$^{-1}$ (Hatzes et al. 2010).
This was consistent with the value of $K$ =  5.27 $\pm$ 0.81 m\,s$^{-1}$
determined using an entirely different filtering approach that  did not rely on
periodic functions  (Hatzes et al. 2011). On the other hand, the
initial $K$ amplitude derived from harmonic analysis was 1.9 m\,s$^{-1}$, or almost a factor
of three lower than the final value (Queloz et al. 2009). Furthermore, the harmonic analysis failed to detect the
presence of a third planetary companion, CoRoT-7d (Hatzes et al. 2010).

In pre-whitening process one usually picks the highest peak in the Fourier amplitude spectrum
of the time series. A least squares sine fit is made to the data to determine
the optimal frequency, amplitude, and phase. This fit is then subtracted from
the data which also removes (or at least minimizes) the effects  of alias peaks
caused by this signal. A Fourier analysis on the residuals then finds the next
dominant peak and the process is continued until one reaches the noise level
of the amplitude spectrum. Each process thus ``whitens'' the data in frequency space
for the next step.  The resulting frequencies that are found should represent
the dominant frequency components of the time series. In our case we use
the sum of the pre-whitened components we have found to provide us with
a fit to the activity variations.

The Fourier components one derives (amplitude, frequency, and phase) depend on the length
of the time series and the presence of time gaps (sampling window). Some
of the effects of these can be explored by analyzing first the complete data set, and then
subsets divided into epochs of the measurements. In doing so one may  derive slightly different
Fourier components in fitting the underlying activity variations.
A robust signal should be relatively insensitive
as to how we pre-whiten the data.

\subsubsection{Fourier Component Analysis of the  Full Data Set}

\begin{figure*}
\plotone{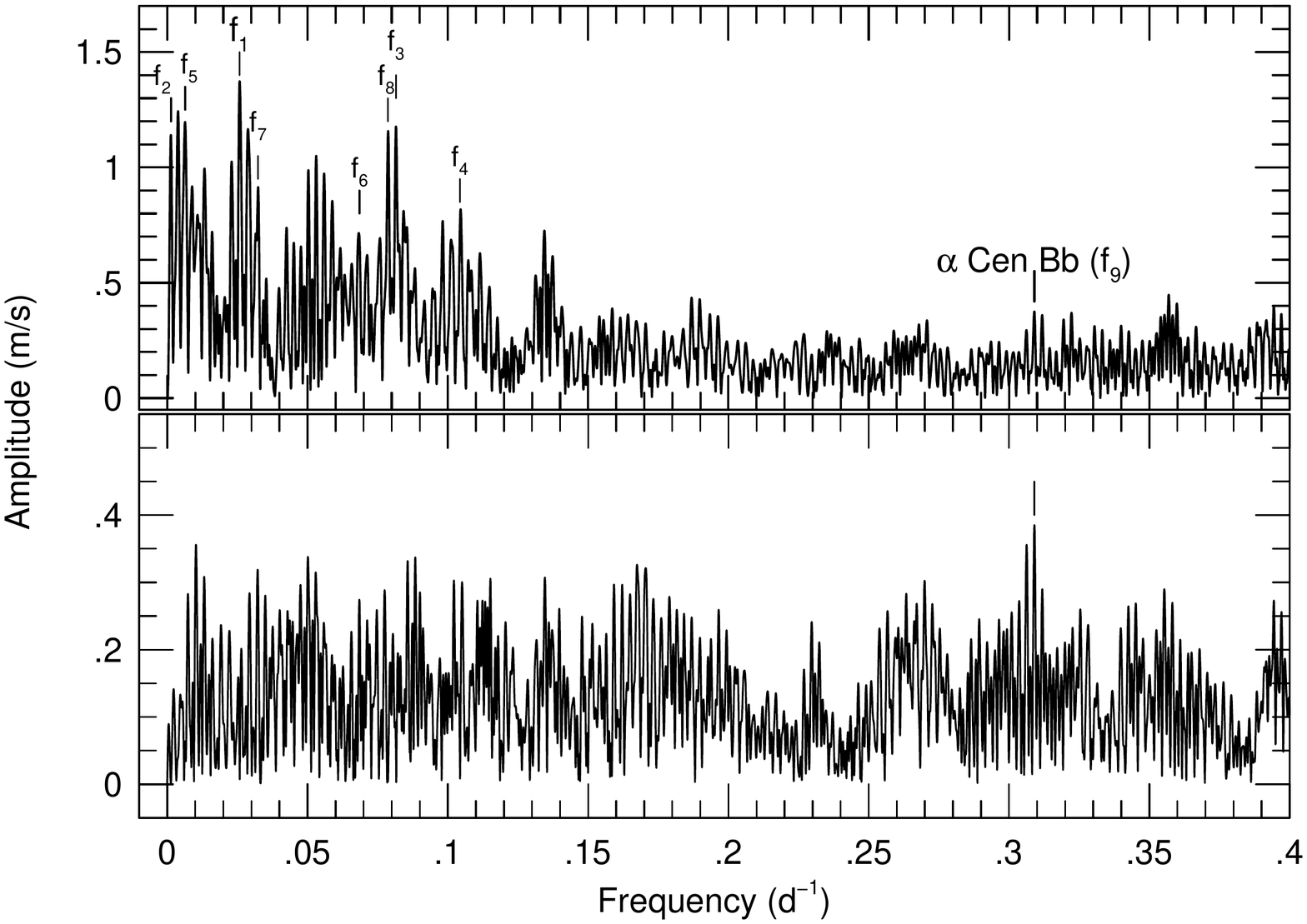}
\figcaption[]{(Top) The DFT amplitude spectrum of the full RV data. The marked
frequencies indicate those found and removed by the pre-whitening process (Table 2).
The orbital frequency of $\alpha$ Cen Bb ($f_9$ from Table 2) is also shown. 
(Bottom) The final pre-whitened amplitude spectrum with only $f_9$ remaining.
\label{prewhite}}
\end{figure*}

The pre-whitening procedure was applied to the full RV data using the program
{\it Period04} (Lenz \& Breger 2004).  For clarity we only show the
DFT of the un-whitened data (top panel Figure~\ref{prewhite}) rather than
each step of the process. 
In the figure we have marked the frequencies found by the pre-whitening process.
Note that removing a peak also removes its alias in Fourier spectrum whose amplitude may be
higher than a real peak.
The amplitude spectrum is dominated by a forest of peaks in the frequency
range 0 $<$ $\nu$ $<$ 0.1\,d$^{-1}$. These may be  due to the rotational frequency of
the star, its harmonics, as well as the spectral window (sampling).
The highest peaks corresponds to a frequency $\nu$ = 0.025\,d$^{-1}$ 
(period, $P$ = 38 d) which is interpreted as the rotational frequency of the star. This frequency
has an amplitude of $\approx$ 1.5 m\,s$^{-1}$ which is about
three times larger than the $K$-amplitude
of the purported planet. Note that the orbital frequency at $\nu$ = 0.309\,d$^{-1}$ is hardly visible
in the original un-whitened data. 

Table 2 lists the frequencies, corresponding periods,  amplitudes, and phases for 
all signals found in the RV time series. 
Note that amplitudes found in Table 2 may differ 
slightly to those
seen in Figure~\ref{prewhite} due to removal of peaks and their aliases.
The lower panel of Figure~\ref{prewhite} shows the final pre-whitened amplitude
spectrum with only the peak near the orbital frequency of $\alpha$ Cen Bb ($f_9$)
remaining. 
Note that in fitting the data simultaneously with all frequencies a slightly
higher amplitude ($\approx$ 1.9 m\,s$^{-1}$) for the rotational frequency
results when compared to the initial DFT. Removing all frequencies in Table
2 from the RV data results in a standard deviation of 1.17 m\,s$^{-1}$.
Interestingly, this value of $\sigma$ is consistent with the rms scatter of Epoch 1 for which D2012
noted was a time when the star was relatively inactive and thus should have a lower
RV jitter. We thus use $\sigma$ = 1.2 m\,s$^{-1}$ as our  ``worse case'' estimate of the RV error.

The low frequency component ($f_2$) has a period, 763\,d, that is shorter
than the time span of the observations of 1230\,d.  This is most likely
due to some variation of the activity, but we cannot exclude with 
certainty that it may be caused by slight differences between the parabolic fit and the true
Keplerian orbit. Unfortunately due to the long period this orbit is poorly known.
Because the period is over a factor of 200 greater than the planet orbital
period this should not effect the subsequent analysis, particularly for the local trend
fitting analysis (see below).
All errors are uncorrelated errors from {\it Period04}; 
correlated errors are most
certainly larger. Many frequencies can be identified with the rotational
frequency, $\nu_{rot}$, or its harmonics: $f_1$ = $\nu_{rot}$,
$f_3$, $f_8$ $\approx$  3$\nu_{rot}$, $f_4$ = 4$\nu_{rot}$, and  $f_6$ $\approx$ 2$\nu_{rot}$. 
This gives some justification to the harmonic analysis used by
D2012.
Note that the last entry ($f_9$) in the table
corresponds to the orbital frequency of $\alpha$ Cen Bb. The pre-whitened
period ($P$ = 3.2356 $\pm$ 0.0001 d)  is identical  to the value
$P$ = 3.2357 $\pm$ 0.0008 d found by D2012.
The amplitude,
$K$ = 0.40 $\pm$ 0.08 m\,s$^{-1}$, is bit lower, but still consistent within the errors
to the value of $K$ = 0.51 $\pm$ 0.04 m\,s$^{-1}$ from D2012. The implied companion
mass from the pre-whitened amplitude is $m$ = 0.89 $\pm$ 0.18 M$_\oplus$.

\begin{figure}
\plotone{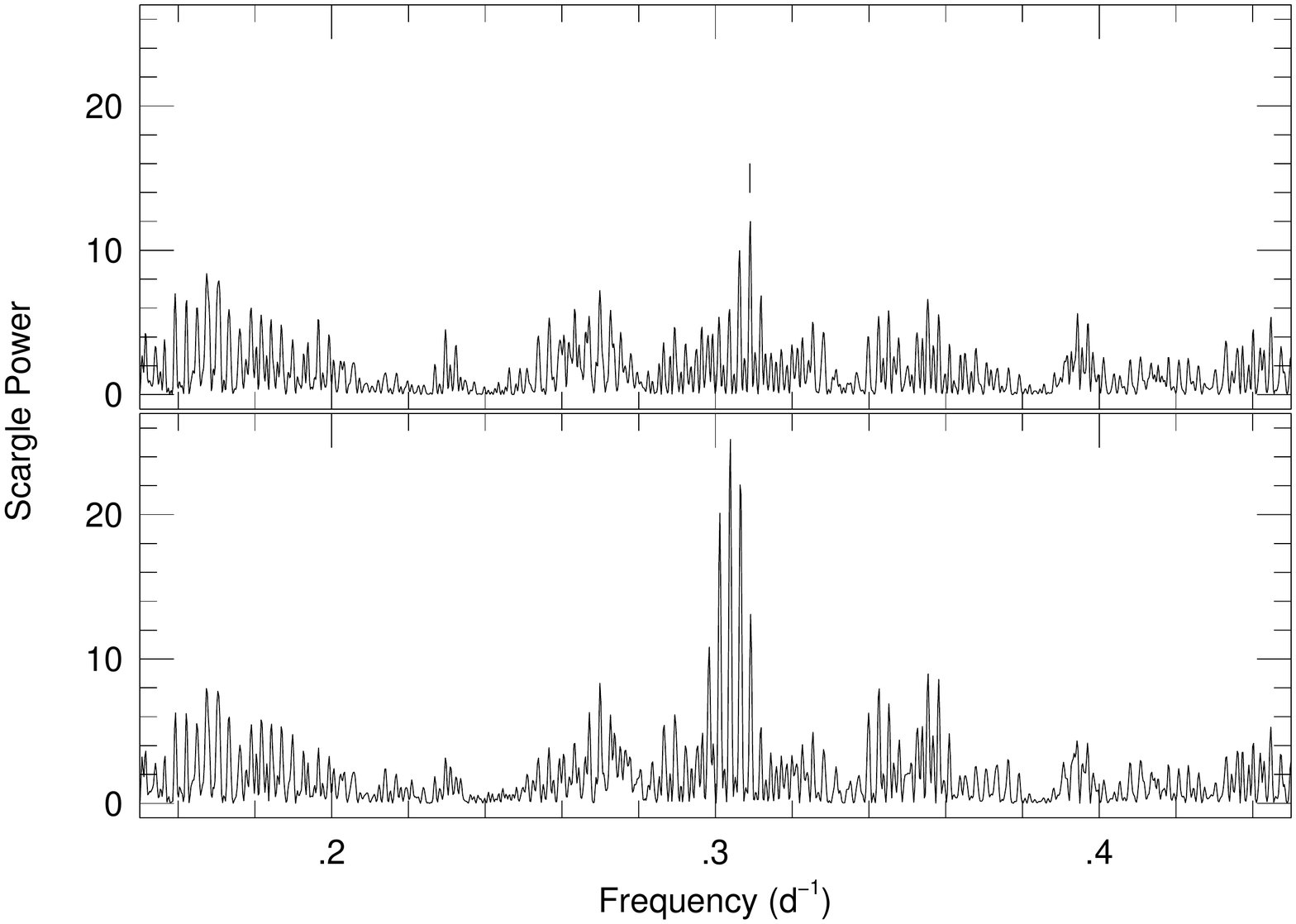}
\figcaption[]{(Top) The Scargle periodogram of the residual RVs of $\alpha$ Cen
after removing the first eight frequencies in Table 1. The vertical line marks the orbital
frequency of $\alpha$ Cen Bb. (Bottom) The periodogram
of the RV residuals with a simulated planet signal ($\nu$ = 0.303\,d$^{-1}$ = 3.29\,days
$K$ = 0.5 m\,s$^{-1}$)
inserted into the data prior to pre-whitening.  
\label{scargle}}
\end{figure}

The statistical significance of the 3.24\,{\rm d} signal was
assessed with a Scargle periodogram analysis (top panel of Fig.~\ref{scargle})
of the residual
RV data produced after removing the first eight frequencies listed in Table 1. 
The peak at the planet orbital frequency of 
0.309\,d$^{-1}$ ($P$ = 3.24 d) appears to be significant due
to its high Scargle power ($z$ $\approx$ 12). The  false alarm probability
(FAP) of this peak was assessed using the bootstrap randomization process. The residual
RV values were randomly shuffled while
keeping the time values fixed. The highest peak in the Scargle periodogram
in the frequency range 0.0001 $<$ $\nu$ $<$ 0.5\,d$^{-1}$  was found for each random data set.
The number of instances where the shuffled data produced power higher than the observed power
provided a measure of the FAP.
The resulting FAP was $\approx$ 0.004 .   (Note that all FAP values given below are 
the result of bootstrap analyses performed with 200\,000 shuffles.)

It is difficult to assess the significance of a periodic signal in the presence of other signals
(in this case at least eight) that have been filtered from the data. 
The FAP calculated 
with a bootstrap depends on the  scatter in the data.
After removing a periodic
signal the amount of scatter in the data is reduced and a much lower FAP may thus
result. However, one cannot be sure that a peak in the amplitude spectrum is a true signal and should
be removed (e.g., rotation), or a noise peak that  should remain in the
data when performing a FAP analysis. 
The bootstrap analysis may produce an unrealistically
low FAP simply  because you have ``cleaned'' the data by 
lowering the noise floor in Fourier space. An alternative approach is to use the
unfiltered amplitude spectrum itself. Kuschnig et al. (1997) established that peaks in the amplitude spectrum
that have a height 3.6 times the surrounding noise level corresponds to
a FAP $\approx$ 1\%.

{\it Period04} calculates a noise level at the orbital frequency of the planet of 0.23 m\,s$^{-1}$.
The amplitude of the planet signal is $\approx$ 0.4 m\,s$^{-1}$
which is only a factor of 1.7 above the noise level.
This corresponds to a FAP of $\approx$ 100 \%. If one
uses the pre-whitened amplitude spectrum with all frequencies in Table 2 removed except for
the planet orbital frequency the noise level is 0.12 m\,s$^{-1}$.
This is the same value as simply taking the mean amplitude of peaks over 
the interval 0.25\,d$^{-1}$ $<$ $\nu$ $<$ 0.35\,d$^{-1}$.
This amplitude is 3.3 times the noise level which corresponds to a FAP $\approx$ 5 \%.
The noise level of the DFT amplitude spectrum thus indicates a 
FAP of approximately a few percent, but with a
large uncertainty. Given the complexity of the RV
variations it is difficult
to assess an accurate false alarm probability.

The efficacy of the Fourier procedure was tested on simulated data. First, the 
3.236-\,day period Fourier component 
(last entry in Table 2) was removed from the HARPS RV data on the assumption that this 
signal is real.
The orbital solution of D2012 was then added back into the data.
 We also added synthetic planet signals at slightly
different orbital periods. 
For these other simulations we used 
the  RV data {\it without} removing the 3.236\,d component on the assumption that this is
due purely to noise in the data.
In this way all simulations preserved the noise characteristics
of the real data.
As an example, the lower panel in  Figure~\ref{scargle}  shows 
a   Scargle periodogram of the residual
RV data with an artificial planet inserted prior to filtering the
data.  The planet signal was taken as a simple sine wave with a period of 3.29\,d
(slightly different from the period of $\alpha$ Cen Bb) and an amplitude of 0.5 m\,s$^{-1}$.

Table 3 summarizes the results of the Fourier filtering of the simulated data.  The third
entry is the simulation using the orbital parameters from D2012.
The table lists the input period of the sine
function, $P_{in}$,
the input amplitude, $K_{in}$, the period recovered by the
pre-whitening procedure, $P_{out}$, the output amplitude $K_{out}$, and the FAP of the
detected signal computed using a bootstrap.
In all cases the input period and amplitude
are recovered well and at 
high level of significance.

\subsubsection{Fourier Component Analysis of the Individual  Epochs}

The pre-whitening procedure was then applied to the individual epochs in Table 1.
In cases where you have periodic signals that are evolving with time (e.g.
the birth, decay, evolution, and migration of surface spots) pre-whitening
of a long time series with large gaps may produce poorer results. A much
better fit to the activity could be obtained by using data covering a shorter time interval.
Furthermore, since one is deriving a different set of sine functions for filtering the
data, this is an independent check on how robust the signal that was
found when analyzing the full data set is.

\begin{figure}
\plotone{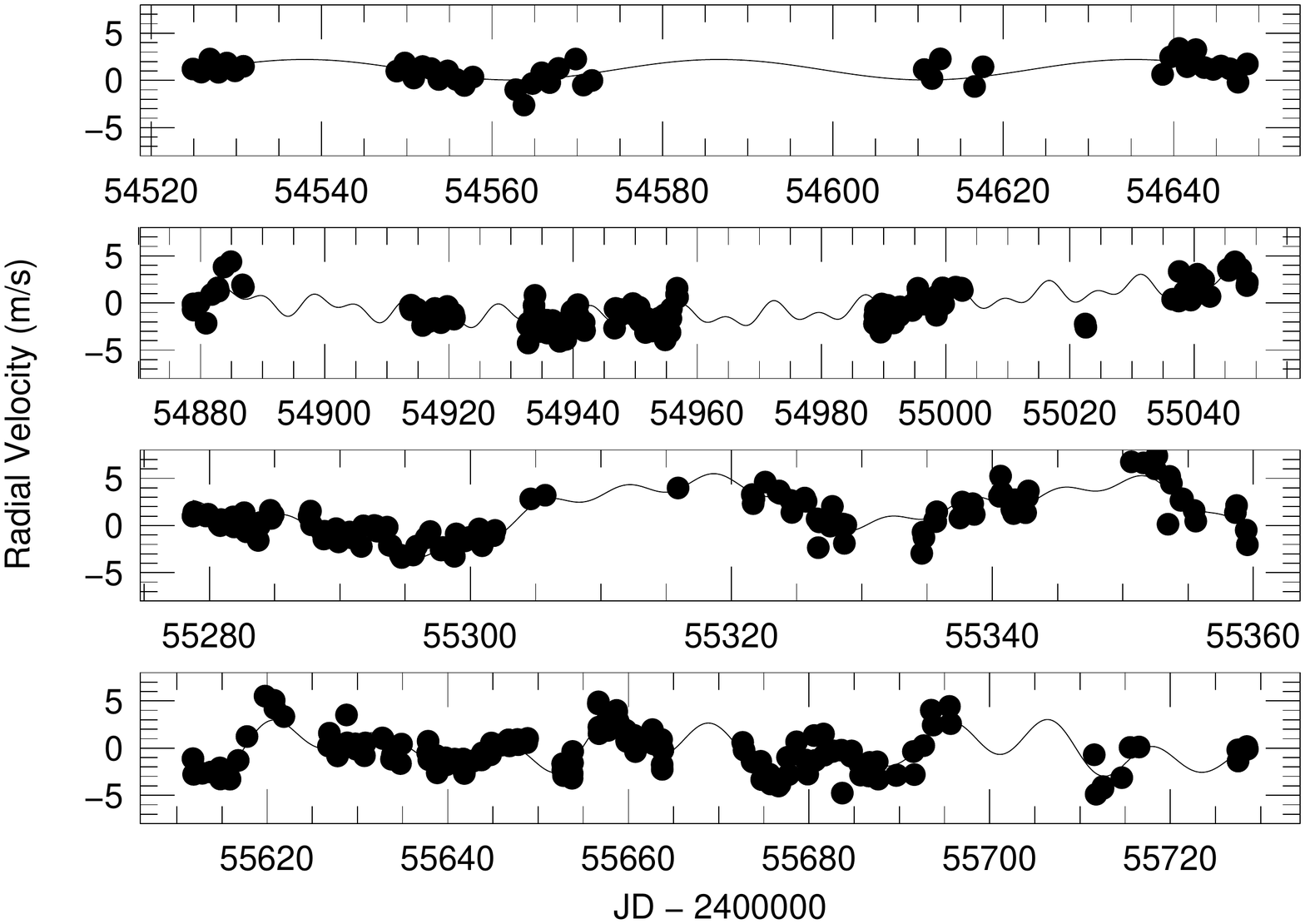}
\figcaption[]{The underlying activity signal for the four epochs computed using the
pre-whitening process separately for each epoch.
\label{epoch}}
\end{figure}

\begin{figure}
\plotone{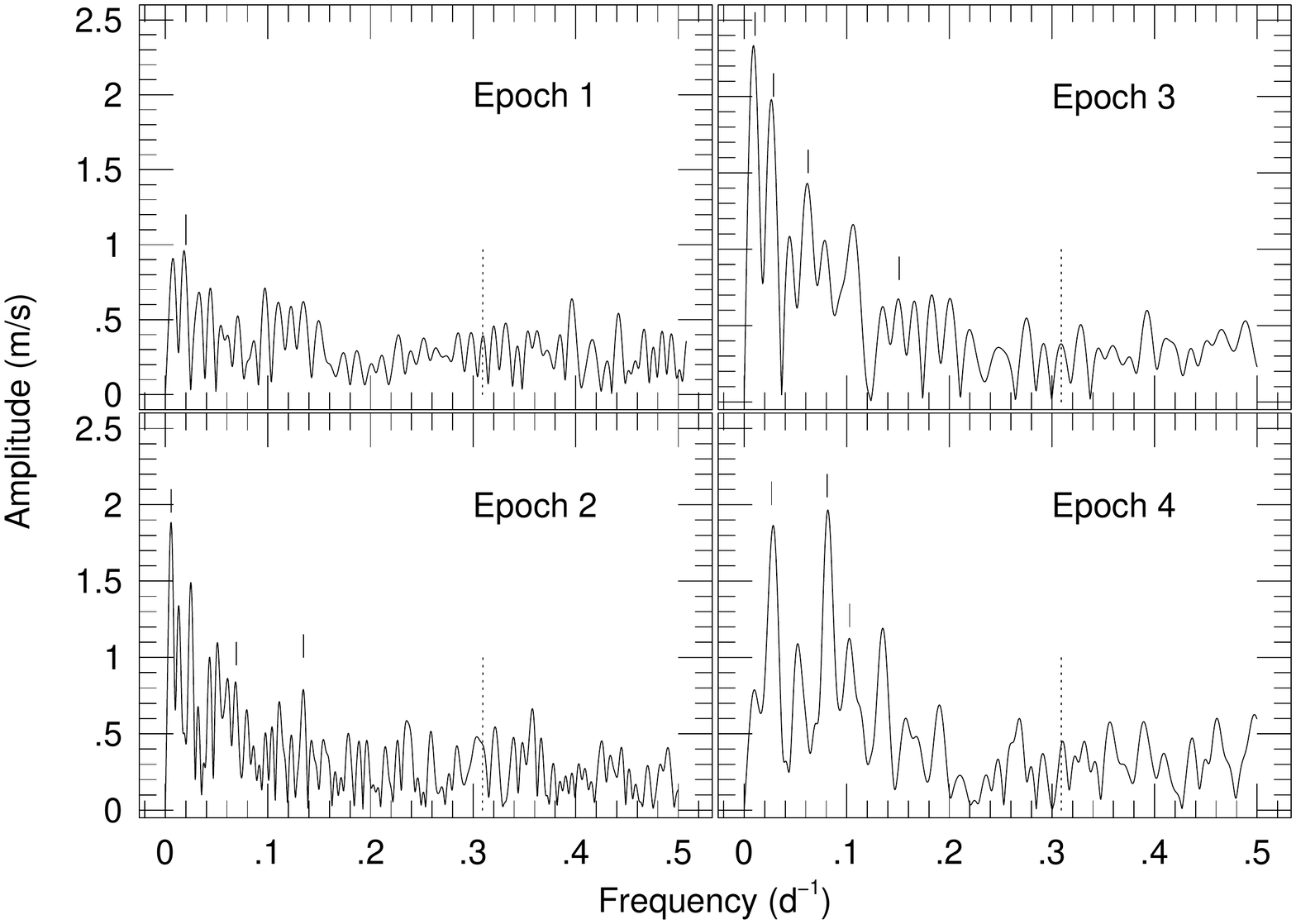}
\figcaption[]{The DFT for the four epochs of RV measurements.
The vertical lines mark the frequencies removed via the pre-whitening process and
the dashed vertical line is the planet orbital frequency.
\label{epochdft}}
\end{figure}

Figure~\ref{epoch} shows the fit to the activity variations
to each of the epochs and Table 4 lists the sine parameters.  The first subscript
in the frequency identification (ID) refers to the epoch number from Table 2.
The DFTs of the RVs from the individual epochs are shown in Figure~\ref{epochdft}. 
In the figure we have marked the frequencies found by pre-whitening the data.
In Epoch 2 the second dominant
peak was found at 0.356\,d$^{-1}$ ($P$ = 2.8\,d). This was not removed since it has a frequency near
that of the planet orbital frequency. 

\begin{figure}
\plotone{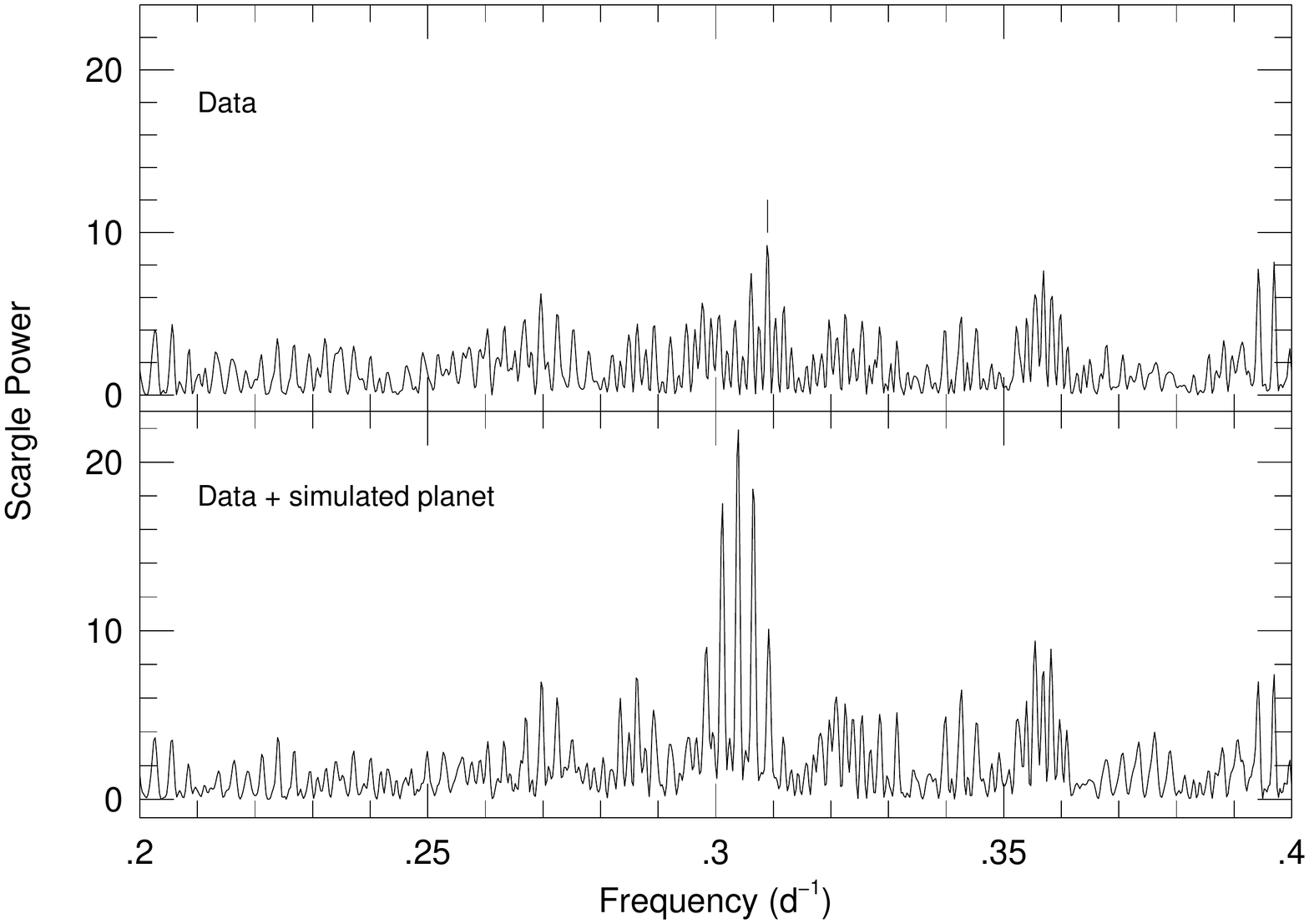}
\figcaption[]{(Top) The Scargle periodogram of the pre-whitened epoch 
RV measurements for $\alpha$ Cen B. The vertical line marks the orbital
frequency of $\alpha$ Cen Bb.
(Bottom) The Scargle periodogram
of the pre-whitened epoch
RV measurements for $\alpha$ Cen B but with the addition of a simulated planet
having $P$ = 3.29\,d ($\nu$ = 0.303\,d$^{-1}$) and  $K$ = 0.5 m\,s$^{-1}$.
\label{epochscargle}}
\end{figure}

The upper  panel of Figure~\ref{epochscargle} shows the periodogram of the total RV residuals
from all epochs.
The epoch pre-whitened RV residuals have a standard deviation of $\sigma$ = 
1.16 m\,s$^{-1}$
with the planet signal present. 
One can still see a peak at the orbital frequency of the planet, but the power is reduced
from that found by the analysis on the full data set. A bootstrap analysis yields
a FAP of only 0.07 for this peak.  Removing the planet signal reduces the standard
deviation slightly to $\sigma$ = 1.13 m\,s$^{-1}$.

The epoch pre-whitening was also tested on simulated data.
A sine wave ($P$ = 3.29\,d, $\nu$ = 0.304\,d$^{-1}$, $K$ = 0.5 m\,s$^{-1}$) was inserted in the
data prior to pre-whitening. 
The lower panel of Fig.~\ref{epochscargle} shows the Scargle periodogram of the 
residual RV data. The FAP of this signal is $<$ 5 $\times$ 10$^{-6}$ based on a bootstrap
analysis. As another test a sine fit to the 3.24\,day period was made to the full 
data set and removed (i.e. removal of $f_9$ from Table 2) and the orbital solution from D2012
inserted back into the data. The pre-whitening procedure was then applied to the epoch data.
The planet signal was detected with a FAP = 0.01 \%.

The pre-whitening process on the epoch data was able to detect the planet, but with much
reduced significance. One would naively think that since essentially the same technique
is applied to both the full and epoch data that we should arrive at the same answer.
Indeed,  we removed from the data the 3.24-d period found in pre-whitening the full data set
and re-inserted the orbital solution for $\alpha$ Cen Bb from D2012. Pre-whitening of the epoch
data showed significant Scargle power of $z$ $\approx$ 15 which  corresponds to 
FAP $\approx$ 0.03 \%. This is consistent with the full data set pre-whitening: 
$z$ $\approx$ 18, FAP $\approx$ 0.002 \%.
Both approaches to filtering the data detects the planet with a much higher significance
than was found.
The discrepancy 
is the first hint that the planet signal may depend sensitively on 
how the data is filtered. Clearly, an independent, non-Fourier based technique is needed
to help resolve this discrepancy.

\subsection{Local Trend fitting}

Although Fourier pre-whitening is a useful tool for getting a quick result, it has its drawbacks.
Some functions, like a linear trend,
may require a large number of Fourier components (i.e. free parameters) to fit them
when simpler functions with fewer free parameters
such as  low order polynomials could provide a better
fit. 
Furthermore, with multi-sine components, if one uses insufficient Fourier components there can be
mis-match between the true activity variations and the fit. 
The sampling window complicates matters further. All of these may introduce false peaks in the filtered
data, or increase the apparent significance of noise peaks.
It is therefore important 
to check the results of pre-whitening and harmonic analysis
by using alternative filtering techniques that rely less 
on periodic functions and that have fewer free parameters. For a robust signal, different filtering approaches
should produce comparable results as was the case for CoRoT-7b. 

A better filter should exploit the fact
that we know the periods  of interest, namely the orbital period of the planet
as well as the rotation period of the star.  If we can fit the activity
variations over a much shorter time span  the signal should be more 
coherent and stable. We thus should be able to use functions with fewer parameters
that fit better the activity variations  at that particular time as opposed to 
using a global fit that requires more free parameters (sine functions).
Trend filtering is often
employed to remove the stellar variations when searching for transit signals in light
curves (Kov\'acs et al. 2005; Grziwa et al. 2012; Bakos et al. 2013).

The time window for fitting the activity variations is bounded by two limits.
The lower limit is the orbital period of the planet. Filtering out activity variations
over a time span less than this runs the risk 
of suppressing any real RV variations due to the planet.
The upper limit is defined by the rotation period of the star, a time
span over which we consider the activity variations likely to be stable and coherent.

The RV data were  visually inspected and divided into time chunks for the local trend fitting.
The following criteria were used to decide which data went into a specific chunk:
1) the time interval of a chunk, $\Delta T$ should cover as many cycles of the planet orbit as possible
but at least a 
full orbital period, but less than a stellar rotation period:
$P_{orbit}$ $<$ $\Delta T$ $<$ $P_{rot}$. 2) The time series should have good sampling (preferably nightly)
in the interval and with no gaps longer than
a day or two. 
3) In fitting trends the data should be  grouped to avoid large gags in temporal coverage or 
abrupt changes in the long term variations in the chunk. 
4)  If successive measurements
were separated by several days, but seemed to follow the overall trend they were kept
in the analysis. If they showed significant departures from the trend that required more
complicated fitting functions (i.e. a high order polynomial with more inflections), 
they were removed.  
In total 41 data points were removed from the original 
HARPS data. 
In short, the data were grouped in subsets that showed 
smooth variations of the underlying trend. The fit to these
should have different
Fourier components that are far removed in frequency from those
of the planetary signal.

The choice of fitting function for the trend in each chunk was 
determined visually  so that the fit to the underlying
activity variations would have a minimal influence on the shorter periodic
variations of the planet. 
If the data in a chunk
showed no long term variations, the average value was  subtracted. If it showed a linear trend
a least squares linear fit was made. In cases where the chunk trend showed curvature a second  or third
order polynomial were used.  Although we tried to avoid using periodic functions, 
in one chunk the underlying
activity variations could best be fit with a multi-sine component (see below).

Two time intervals  RJD = 54935--54955 and
 RJD = 55672--55692 showed significant periodic variations.
In the first of these intervals the RV data centered on 
RJD = 54939.68 -- 54941.85  showed
an additional linear trend on top of the sinusoidal variations
(look ahead to the lower left panel of Figure~\ref{newpanels}). Since there was a large gap 
of five days with the first group of points in this interval these 12 points
were removed (in the lower left panel of Figure~\ref{newpanels} these
are marked by the bracket).
This  ensured a simpler fitting function with fewer inflections. (In \S3.4 we will see that this one
time chunk can have a large influence on the amplitude spectrum.) 
The time interval RJD = 54935--54955 was thus divided into
two chunks. In the first (chunk \#8) a second order polynomial
was used to fit the trend, and a third order polynomial for the 
next chunk (chunk \#9).

The interval RJD = 55672--55692 had good
sampling and the data looked periodic. Two sine
functions with
$\nu$ = 0.043\,d$^{-1}$, $K$ = 1.2 m\,s$^{-1}$, 
and $\nu$ = 0.08\,d$^{-1}$, $K$ = 2.3 m\,s$^{-1}$, values found by
pre-whitening the data were used to fit the activity
variations.   These data comprised chunk \#20.

\begin{figure*}
\plotone{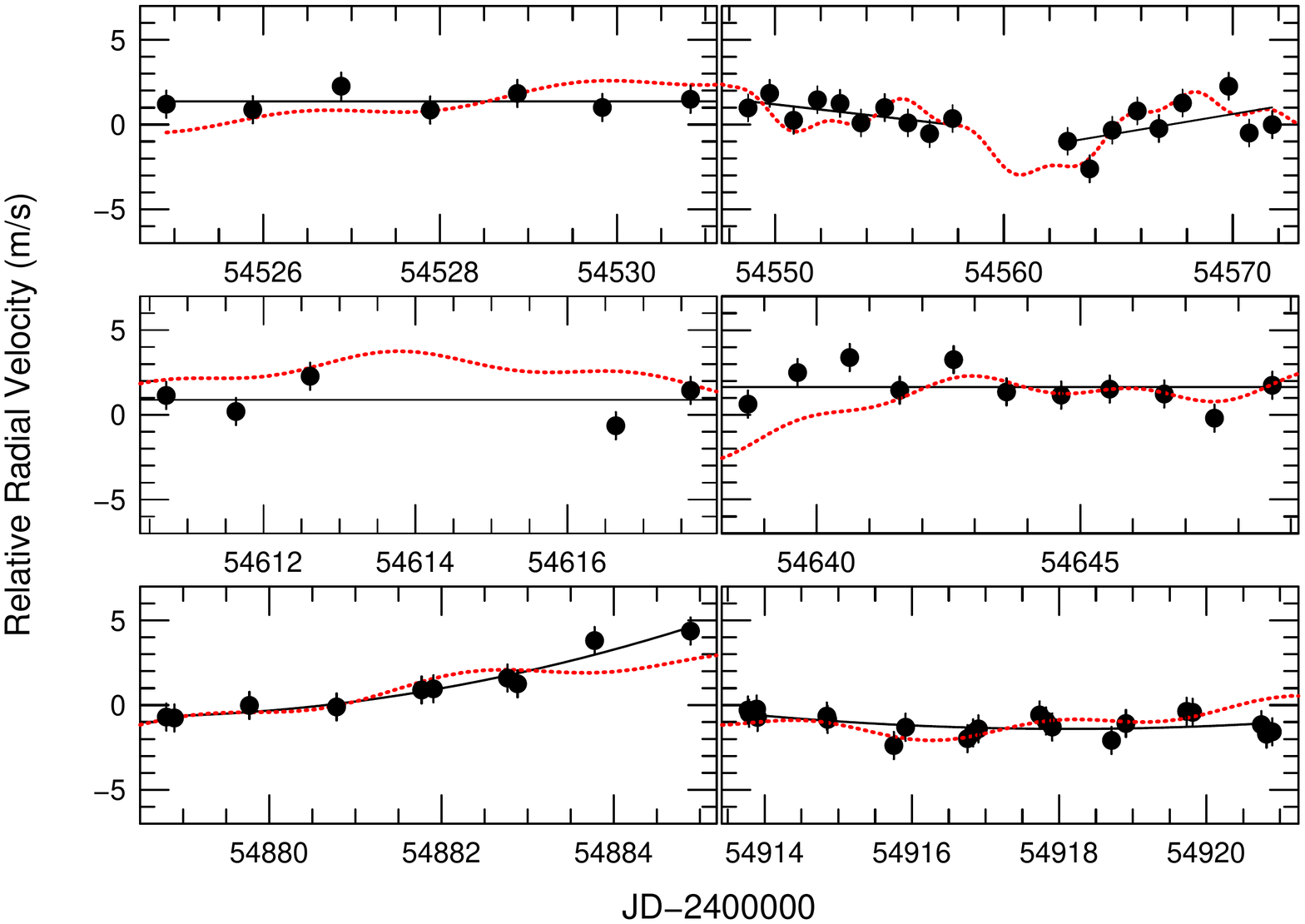}
\figcaption[]{The first six time chunks used for the local trend fitting
of the activity signal. The solid line represents the local fit to the
trend. The dotted red line is the fit to the activity using the pre-whitening process on the 
full data set (Table 2 but without the planet signal). 
\label{panel1}}
\end{figure*}

\begin{figure*}
\plotone{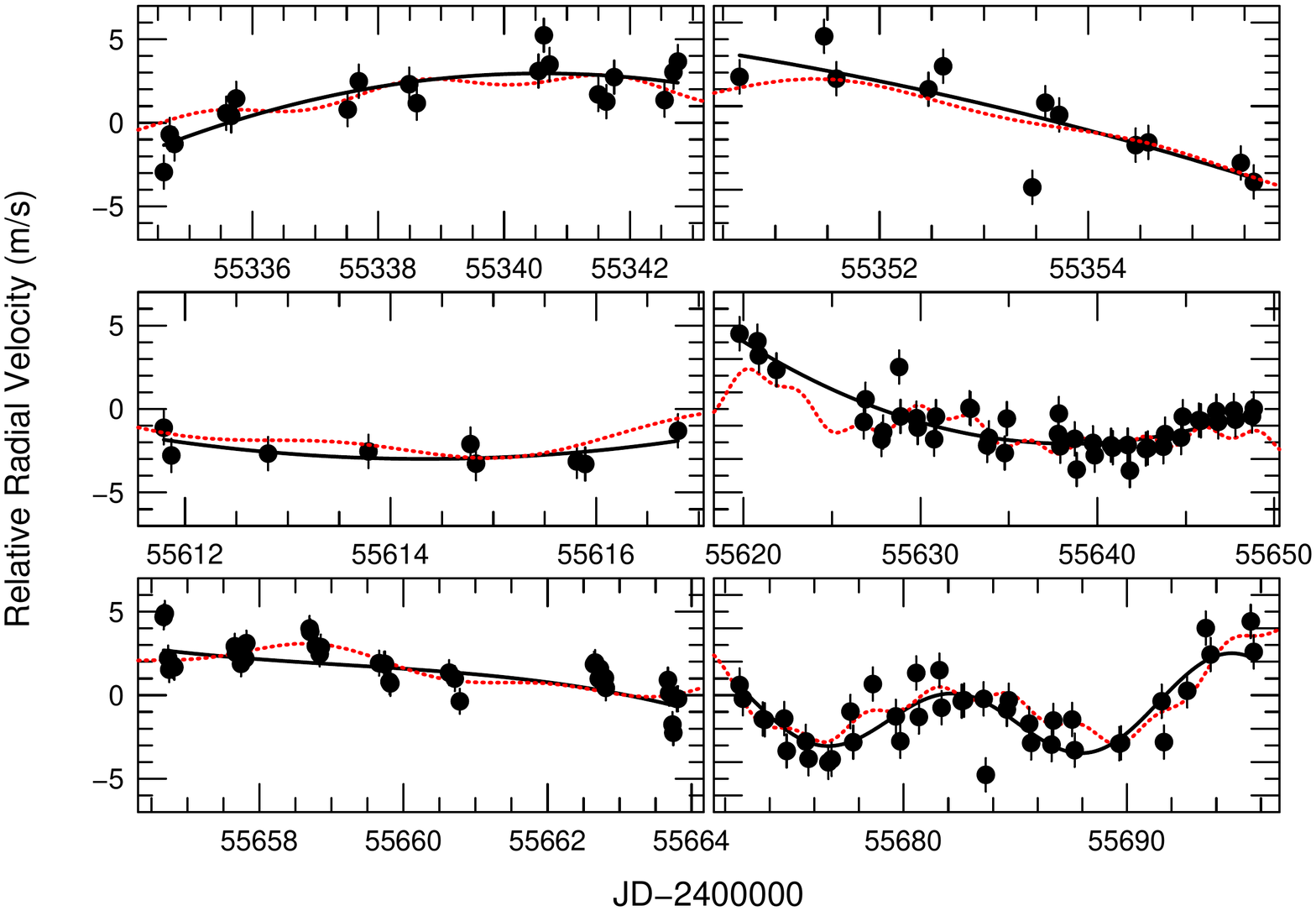}
\figcaption[]{Same as for Fig.~\ref{panel1}, but for the next six time chunks.
\label{panel2}}
\end{figure*}

\begin{figure*}
\plotone{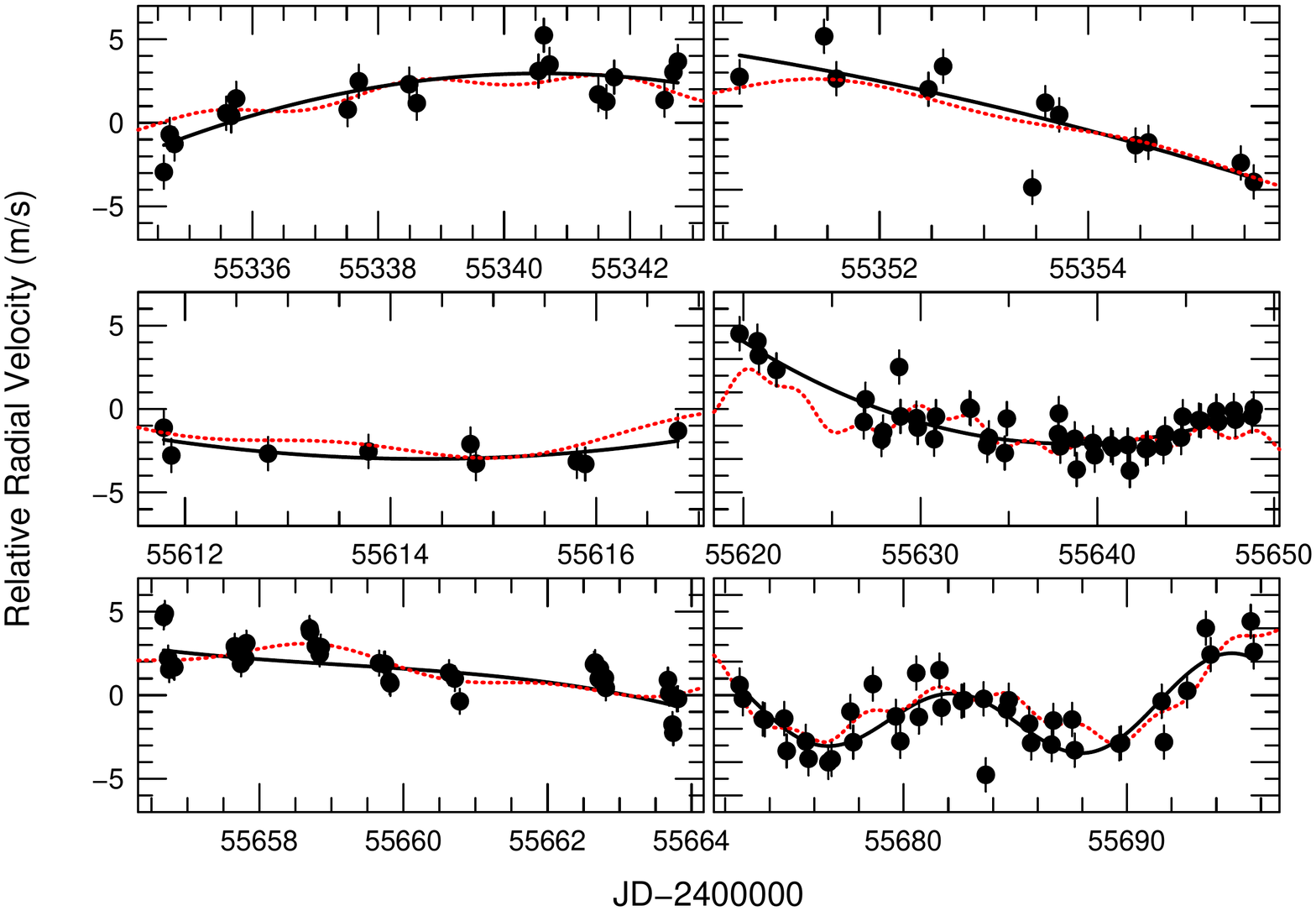}
\figcaption[]{The next six time chunks used for local trend fitting. 
\label{panel3}}
\end{figure*}

Table 5 lists the Julian day of the time chunks, the time span, $\Delta T$, the 
number of planet orbits during this span, $N_{orb}$, the number of data points used in the fit, $N_{data}$,
the fitting function employed (constant = average value subtracted, linear, second or third
order polynomial),
and the rms scatter in the chunk, $\sigma$, after removing the trend but with the planet
signal present.
Figures~\ref{panel1}--\ref{panel3} show the individual time chunks used in the analysis of the
HARPS data. The error bars represent the best case error of  0.8 m\,s$^{-1}$.
The solid lines represent the fit to the underlying
trend in the chunk. As comparison, the fit to the activity using the
sine parameters found by pre-whitening the full data set (Table 2, but without the planet
contribution) is also shown. Note that there are many instances where the LTF
provides a much better fit to the underlying activity variations.

\begin{figure}
\plotone{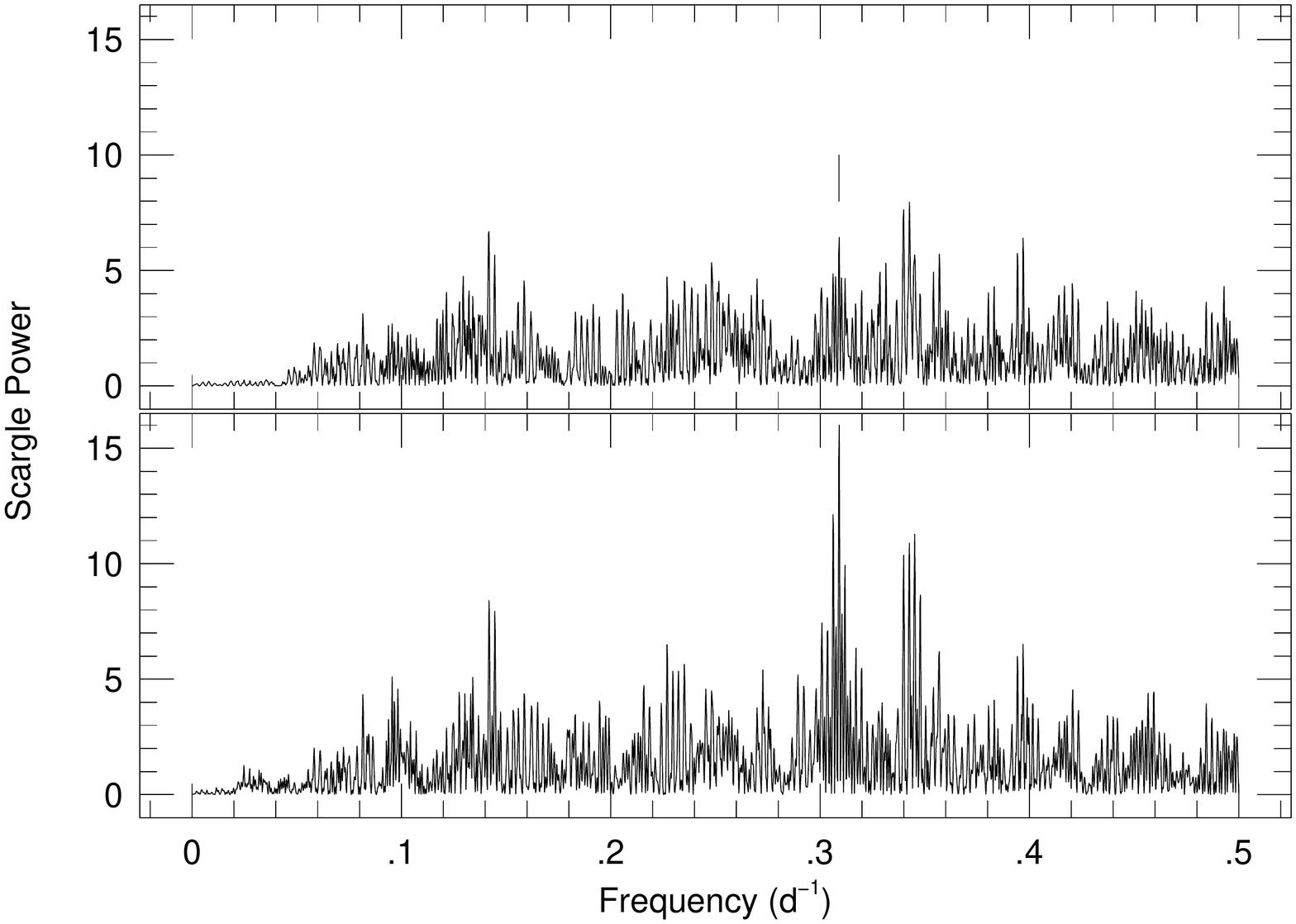}
\figcaption[]{(Top) The periodogram of the RV residuals after removing the local
trends shown in Figures~\ref{panel1}-\ref{panel3} (LTF1). (Bottom)
Periodogram of the local trend fitted RV residuals using the RV data after removing the 
sine component at the planet orbital frequency found by pre-whitening ($f_9$ in Table 2)
and inserting the orbit of $\alpha$ Cen Bb ($P$ = 3.24\,days, $K$ = 0.51 m\,s$^{-1}$).
\label{noplanet}}
\end{figure}

The residual RVs after removing the underlying trend were then combined.
These had a standard deviation of $\sigma$ = 0.94 m\,s$^{-1}$.
The Scargle periodogram (top panel Figure~\ref{noplanet}) shows no significant
peak anywhere in the frequency range $\nu$ = 0 -- 0.5\,d$^{-1}$. 
The peak at the planet orbital frequency is weak and has a FAP of
$\approx$ 0.4 as determined by a bootstrap. As a quick test the Fourier component at
3.24-d ($\nu$ = 0.309 d$^{-1}$, $f_9$ in Table 2) was removed from the full
data set and the orbit of $\alpha$ Cen Bb 
of D2012 added back to the data. The data with the simulated planet were divided into chunks
and trend filtered as the previous data, and calculating new trends for the
data. The Scargle periodogram of the residuals
(lower panel Figure~\ref{noplanet}) shows that the planet signal should have been easily
detected and with a FAP $\approx$ 0.01\%.
{\it The local trend filtering
method does not confirm the presence of a planetary signal at 3.235 d around $\alpha$ Cen B.}

\subsection{Tests of the Local Trend Filtering Procedure}

The local trend filtering (LTF) procedure was tested further to 
see how well it could
recover known signals in simulated data generated in a different ways. 
This was done on three different simulated data sets:

\begin{enumerate}

\item The 3.24\,day planet period was removed from the RV data using the sine function
parameters found by the pre-whitening process. A sine function with the
same amplitude as the planet (0.5 m\,s$^{-1}$) but with a slightly different period,
$P$ = 3.27 d, was inserted back into the data. A different period was employed 
simply to avoid the frequency 
in the amplitude spectrum where signal was removed. This simulation keps the original
noise characteristics of the data.

\item A signal 
with a period of 3.37\,days
and a $K$-amplitude of 0.5 m\,s$^{-1}$
was inserted into the RV data.
A different period to that
of the planet was used because this possible signal is still in the data and we want
to avoid interference between the two. (As
far as LTF is concerned a 3.37\,day period is no different
from a 3.23\,day period.) In this simulation all the noise characteristics of the data
are kept, and with the assumption that the signal at 3.24\,days is also due to noise.

\item A simulated activity signal was generated using the first eight frequencies, amplitudes, and phases
listed in Table 2. (Hereafter we will refer to this simulated
activity signal as ``the activity function''.) This was then sampled in the same way as the data and random
noise with standard deviation, $\sigma$ = 0.8 m\,s$^{-1}$ was added.  The orbit of D2012
was then inserted into the data.
The advantage of this
simulation is that the underlying trend for each chunk may be slightly 
different to the cases above.
The disadvantage is that the noise characteristics,
which are now  Gaussian, may be different than for the real data.

\end{enumerate}

\begin{figure}
\plotone{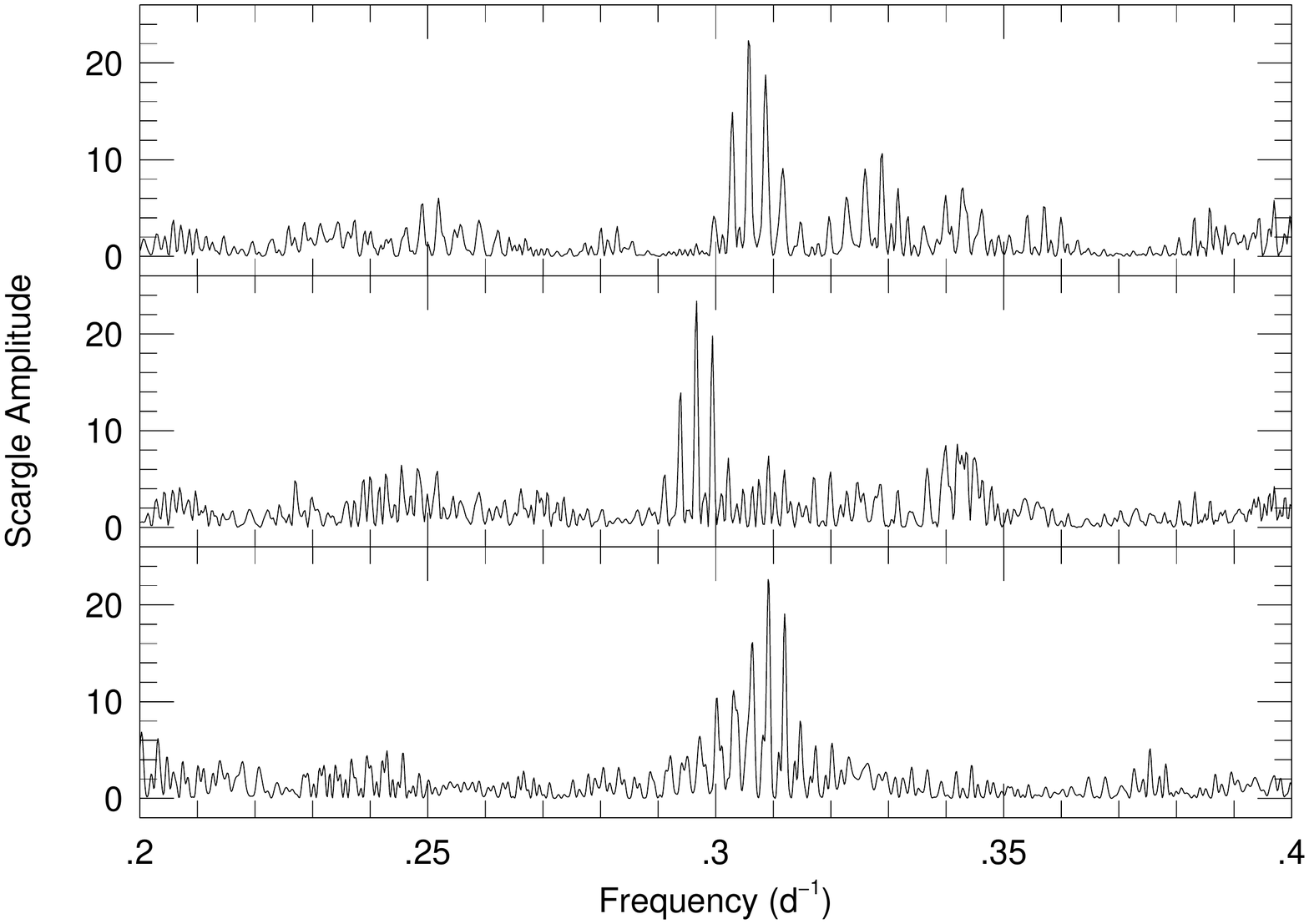}
\figcaption[]{Periodograms of simulated data after applying local  trend fitting.
(Top) Simulation using the actual RV data where the planetary signal was removed
and an artificial  signal with $P$ = 3.27 days, $K$ = 0.5 m\,s$^{-1}$ added back into the data before 
pre-whitening.
(middle) Simulation using the real RV data and a simulated   signal with $P$ = 3.37 d, $K$ = 0.5 m\,s$^{-1}$ added.
(Bottom) Simulated data using an  activity signal consisting of a multi-sine fit
generated with the first eight frequencies in Table 2. This activity signal
has the same temporal sampling as the data and random noise at a level of 0.8 m\,s$^{-1}$.
A planet signal with $P$ = 3.24\,days and $K$ = 0.5 m\,s$^{-1}$ was also added to the data.
In all cases the input signal was recovered at high statistical significant.
\label{ftsim}}
\end{figure}

The LTF technique 
was applied to each of these three artificial  data sets. Although the time span for each
chunk was  kept fixed, the fit to RV data was repeated in each case. Scargle
periodograms of the trend-removed residuals (Figure~\ref{ftsim}) show  that
the local filtering process was able to 
recover the input signal and at a  high significance in all cases.
A bootstrap analysis with 200\,000 shuffles  showed no instance where the periodogram of random data exceeded
the real  periodogram (FAP $<$ 5 $\times$ 10$^{-6}$). 
Note that these signals  were detected at a 
much higher level of significance than 
the original data. Tests using planet orbital variations with a different phase
produced similar results. It appears that 
the filtering process is not suppressing a possible real signal from a planet.

\begin{figure}
\plotone{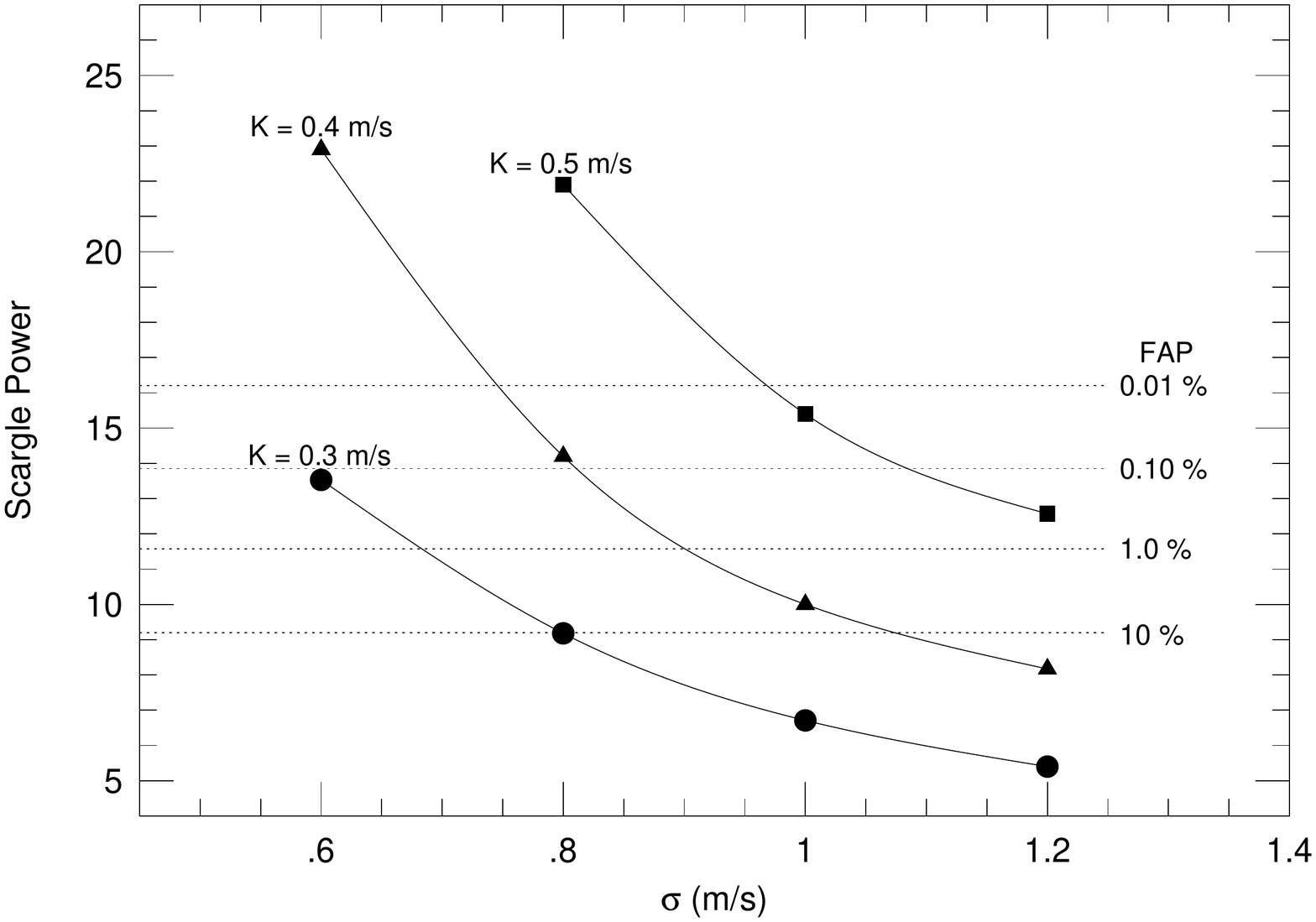}
\figcaption[]{The detected $K$-amplitude for a 3.24-day period planet in the HARPS data
as a function of the rms scatter, $\sigma$, of the data. This is based on simulated data
using the multi-sine component model for the activity and the orbital parameters from D2012,
but with different $K$-values taken from D2012. The ordinate is in Scargle power and the
horizontal lines show the corresponding FAP determined via a bootstrap.
\label{amplimits}}
\end{figure}

The last simulation assumed our best case estimate of the noise. It is of interest to 
explore the detection limits of the LTF procedure  as a function of different noise levels.
Again synthetic data consisting of the activity function
plus the planet orbit parameters of D2012 were used, but this time
with different levels of random noise added.  Figure~\ref{amplimits} shows the Scargle power of
the RV residuals as a function of the standard deviation, $\sigma$, of the random noise.
Shown are simulations for three values of the $K$-amplitude (0.3, 0.4, and 0.5 m\,s$^{-1}$).
If the real measurement error is close the best case value a $K$-amplitude of
$\approx$ 0.35 m\,s$^{-1}$ could be detected with a FAP = 1\%. For a $K$-amplitude of
0.5 m\,s$^{-1}$ the planet could be detected at the 1\% level even for $\sigma$ as high
as $\approx$ 1.4 m\,s$^{-1}$
For planets with
certain orbital periods it should be possible to overcome the activity noise.

\begin{figure*}
\plotone{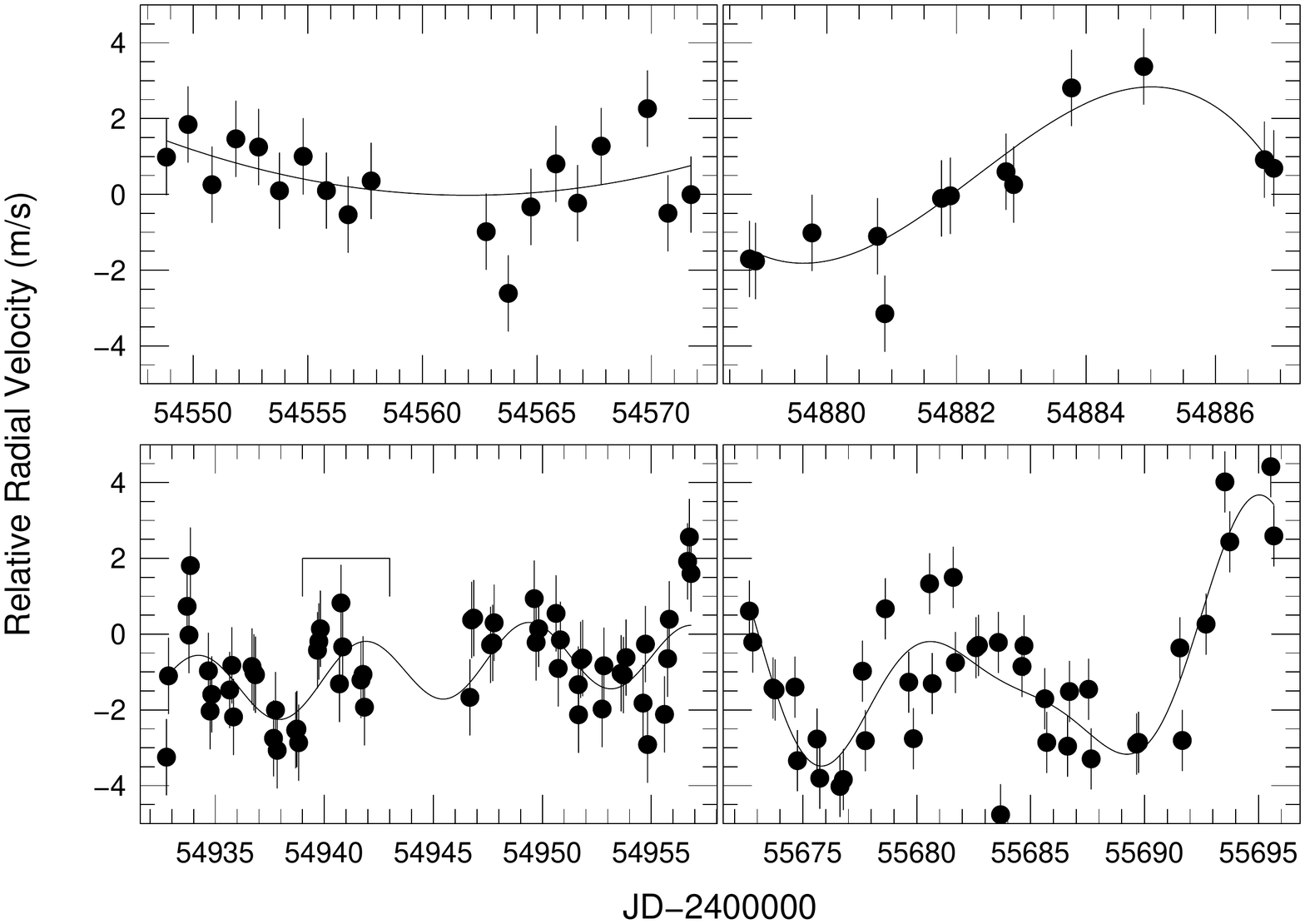}
\figcaption[]{Time chunks and the underlying trend fits used in the second
version of local trend filtering (LTF2). The bracket in the lower left
panel shows points removed in the LTF1 analysis (see text).
\label{newpanels}}
\end{figure*}

The LTF method was also used with different time windows and fitting functions
to explore as in the case of pre-whitening how different filters 
could influence the results. A second trend filtered version
of the data (hereafter LTF2) was made with the following minor differences compared to 
previous version of the trend filtering (hereafter LTF1):

\begin{enumerate}

\item

In LTF1 the data taken during RJD = 54549 -- 54572 
were divided into two chunks because
of a 5-day gap. A linear fit to each was made separately (chunks \#2 and \#3).
In LTF2 a parabola was fit to all the data (upper right panel of Figure~\ref{newpanels})
across the gap.

\item In chunk \#6 of LTF1 a parabola was fit to the data, but after removing the last data points 
that came after a 2-day gap. In LTF2 these points were included, but it forced
one to use a higher order 
polynomial (top right panel of Figure~\ref{newpanels}).

\item 
For measurements made during  RJD = 54933 -- 54956  the data were divided into 
two chunks (chunks \#8 and  \#9) in LFT1 because of 
a five day gap in the time sampling.
In LTF2 data in chunks \#8 and \#9 were combined
and a multi-component sine function was fit to the underlying 
trend throughout the time interval (lower left panel of Figure~\ref{newpanels}).

\item For chunk \#20
an additional sine component was used as determined from pre-whitening
the data (lower right panel Figure~\ref{newpanels})

\end{enumerate}

\begin{figure}
\plotone{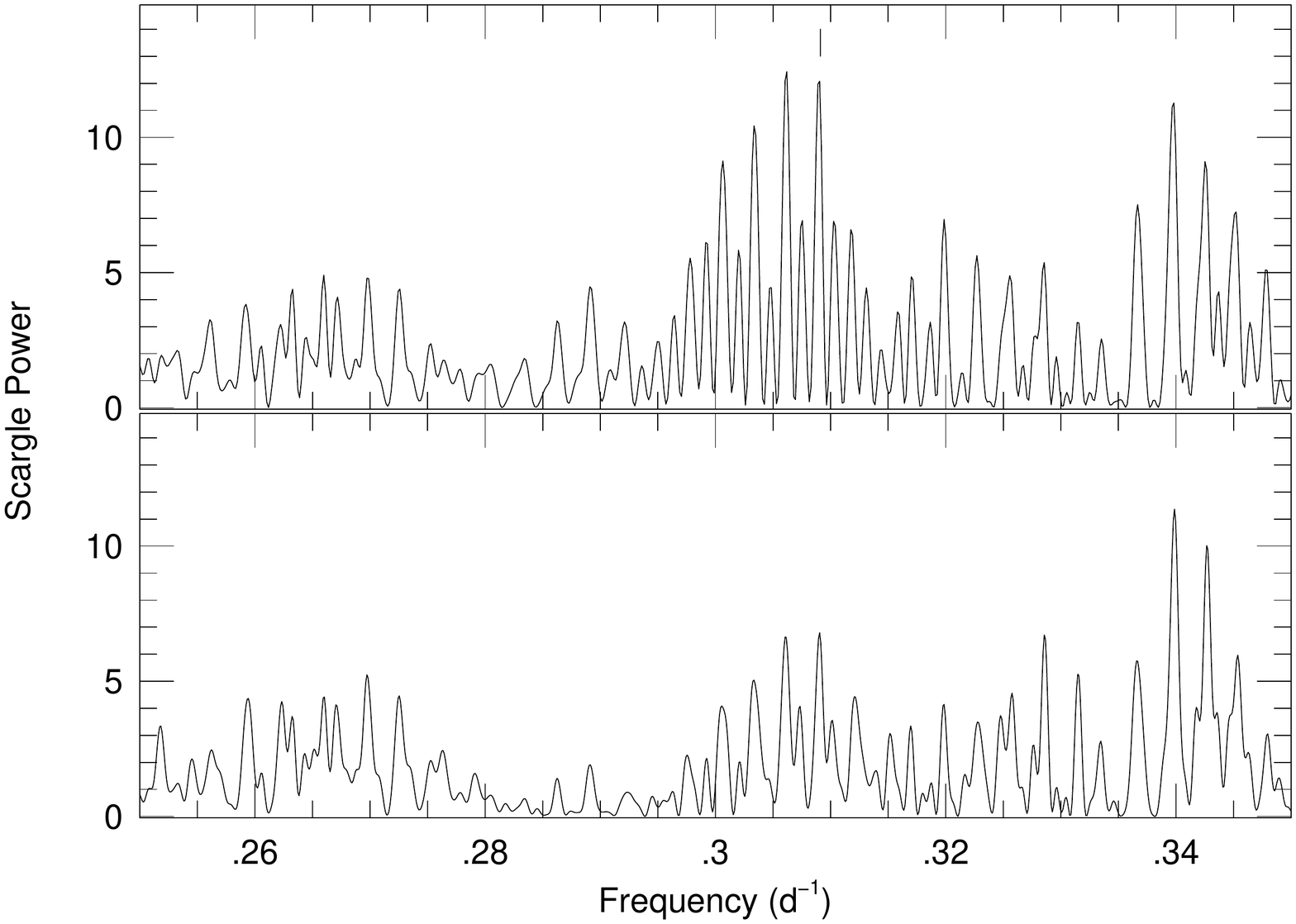}
\figcaption[]{(Top) Periodogram of the RV residuals after applying
the second local trend fitting (LFT2).
The vertical
line marks the orbital frequency of $\alpha$ Cen Bb.  (Bottom) The periodogram
the LFT2 residuals but with different filtering of the data in Chunk8-9.
\label{vers1allft}}
\end{figure}

Figure~\ref{newpanels} shows the new trend fits to these time chunks. 
The total rms scatter of the data (including the other chunks)
after removing the trends is about  1.00 m\,s$^{-1}$, or slightly
worse than for LTF1. For
all the other chunks the same time windows and fitting functions were employed
as per LTF1.
The periodogram of the combined trend-removed residuals 
shows modest power
power (FAP $\approx$ 0.006)
at the frequency near
with the orbital frequency of the planet (top panel of Figure~\ref{vers1allft}).
(Also, the highest peak occurs at  a slightly different frequency, 
$\nu$ = 0.30615\,d$^{-1}$, or $P$ = 3.26\,days).
However, it is highly suspicious that a markedly different
periodogram is obtained when altering slightly the fit to a small subset of the data.
This simulation only reinforces what was found in performing the pre-whitening on the 
full and epoch data sets -- slight differences in the way the RV data is filtered can produce
dramatically different results.

\subsection{The Significance of the LTF2 Detection}

In the course of filtering the activity signal from the individual chunks we discovered
that data in 
the time window JD-2400000 = 54933--54956 (hereafter referred to as
``Chunk8-9'' since it is a combination of chunks \#8 and \#9 in Table 5)  had peculiar frequency
characteristics that may influence the outcome when using different approaches to filtering
the activity signal. A 
Fourier analysis of this chunk revealed a significant signal at 3.3\,days.
Pre-whitening of the data reveals two additional 
frequencies
(Table 6) with frequencies near the rotational frequency, $\nu_{rot}$, and $\approx$ 5$\nu_{rot}$.
The 3.3-day signal is of particular interest 
since this is uncomfortably
close to the planet period. However, it is unlikely that this is due entirely to the planet since its amplitude
is too large by a factor of two.
This signal is significant as
a bootstrap analysis shows that FAP = 
3.5 $\times$ 10$^{-4}$. In producing the residuals from Chunk8-9 that were used in LTF1 the
3.3-d period was kept in the data for the obvious reasons that it nearly coincides with the planetary signal.

As a test we tried different filtering of the data only in Chunk8-9. The first and
last half of the data in the chunk were filtered in different ways (e.g. second, third order polynomials,
or sine functions). Different fits were performed after deleting points after a large
time gap, first and last points from the data, data showing large variations with respect
to adjacent values, etc. Ten different versions of filtering the chunk
were tried and in all cases no more than 20 points were removed. The residuals
were then added to the rest of LTF2 residuals. The Scargle power of the total residuals from
all chunks ranged from
as low as $z$ = 7.0 to as high as $z$ = 11.4. The average power was 8.8 $\pm$ 1.3. This
corresponds to a range in FAP of 3 -- 30\%.  We also took the original LTF2 residuals
from Chunk8-9 and replaced  the corresponding time values in LTF1 and this alone boosted  the
Scargle power at the planet orbital frequency from $z$ = 6.4 (FAP = 0.4) to
 $z$ = 9.75 (FAP = 0.05).

In most  cases the highest power in the periodogram in
the frequency range 0.25\,d$^{-1}$ $<$  $\nu$ $<$ 0.35\,d$^{-1}$ was not at the planet 
orbital frequency, but rather $P$ = 2.94\,days ($\nu$ = 0.3397\,d$^{-1}$).
The lower panel of Figure~\ref{vers1allft}) shows one such filtered version of Chunk8-9
where the dominant peaks are not coincident with the planet orbital frequency.

\begin{figure}
\plotone{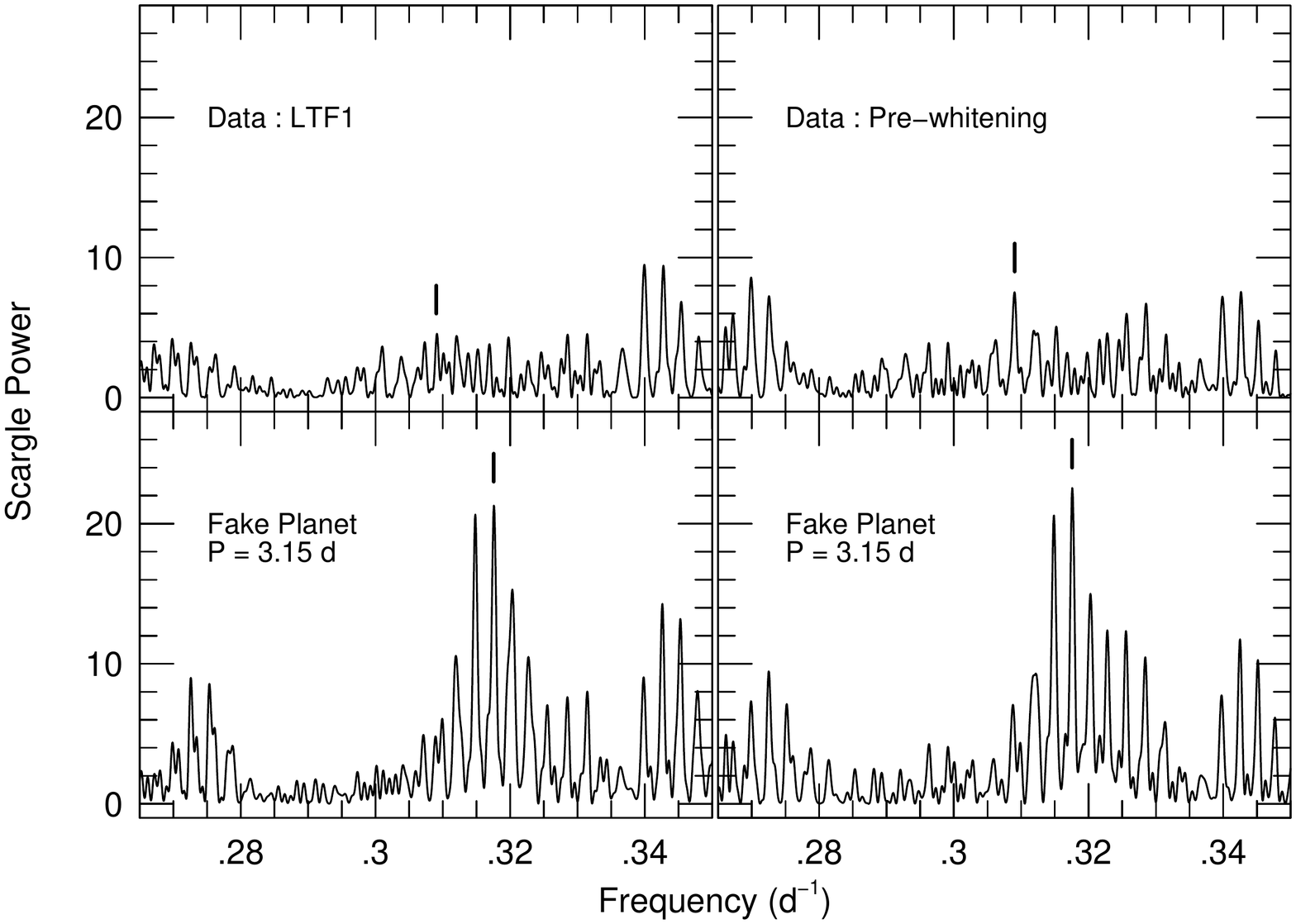}
\figcaption[]{(Upper left panel) Scargle periodograms of the activity filtered
RV data LTF1, but with Chunk8-9 removed. (Lower left panel) The Scargle
periodogram of LTF1 filtered RV data minus Chunk8-9, but with an artificial
planet signal ($P$ = 3.15 days; $\nu$ = 0.3175\,d$^{-1}$, $K$ =
0.5 m\,s$^{-1}$) inserted into the data prior to filtering.
(Upper right) Scargle periodogram of the RV data without Chunk8-9
after filtering the activity signal with pre-whitening.
(Lower right) Same as for the lower left panel (data with an artificial planet
signal added) but for the pre-whitening procedure. The vertical line marks
the location of the orbital frequency of $\alpha$ Cen Bb (top panels)
or the simulated planet signal (lower panels).
\label{c89test}}
\end{figure}

We checked whether the planet signal could be detected in 
the subset RV measurements without data from 
Chunk8-9. Local trend filtering (LTF1) clearly shows no significant signal in this subset data
(top left  Figure~\ref{c89test}). However, a 3.15-day periodic
signal ($K$ = 0.5 m\,s$^{-1}$) that was inserted
into the data was found after applying LTF1 even without Chunk8-9. 
The same results were found when pre-whitening the RV data
without Chunk8-9.
A weak, but insignificant peak is found at the planet orbital
frequency (top right panel Figure~\ref{c89test}). Applying the pre-whitening
to the data with the artificial  planet ($P$ = 3.15 days) can recover the input signal
at a much higher significance (lower right panel in Figure~\ref{c89test}). In
summary, both pre-whitening and LTF methods should have been able to detect the
planetary signal of $\alpha$ Cen Bb even without the data from Chunk8-9, but they do not.

\begin{figure}
\plotone{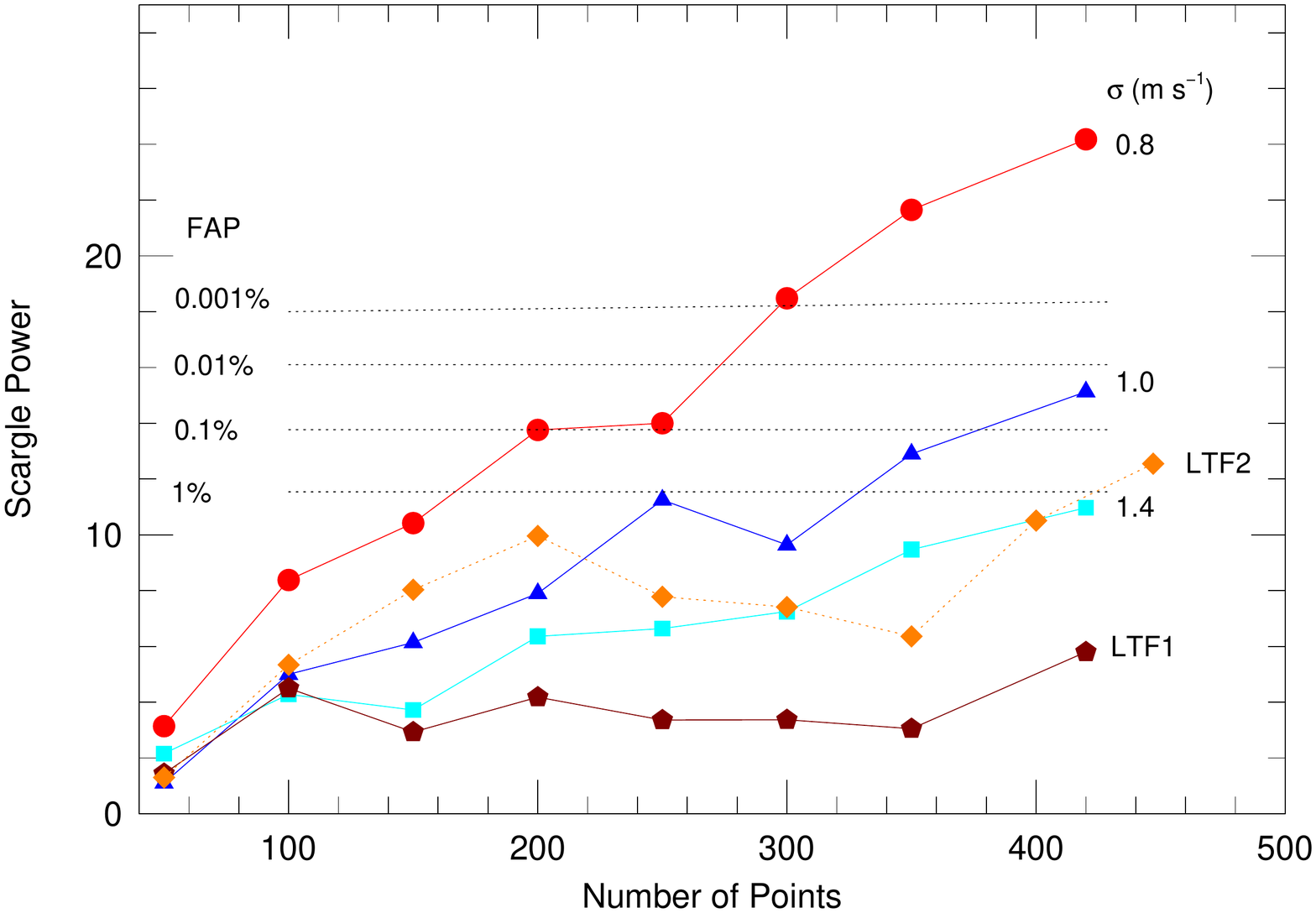}
\figcaption[]{The Scargle power
of the planet signal in the residual data from LTF1 (pentagons) and LTF2
(diamonds) as a function of number of data points.
The same is shown for simulated data taking the activity function,
a synthetic planet signal, and applying local trend fitting (time windows of LTF1).
Noise has been added at three levels: $\sigma$ = 0.8 m\,s$^{-1}$ (dots),
1.0 m\,s$^{-1}$ (triangles), and 1.4 m\,s$^{-1}$ (squares).
\label{fap}}
\end{figure}

The behavior of the Scargle power as a function of the number of data points is a good
way to assess the significance of a real periodic signal (see Hatzes \& Mkrtichian 2004).
A real signal should have Scargle power that increases in an expected way
as you add more data.
The behavior of the statistical significance for the complete data 
was also inconsistent with the expectations of a real signal. Figure~\ref{fap}
shows the power at the orbital frequency of the planet as a function of the number
of data points, $N$, using the residuals from local trend
fitting  and adding data sequentially
in chronological order. We show the results for LTF1 and LTF2.

The figure also shows three simulated data sets.
For these  we 
added the orbit of $\alpha$ Cen Bb  to the activity function generated from  the sine
components found by the pre-whitening of the
individual epochs.
 The total RV curve was then sampled in the same
way as the data and three different levels of random noise were added ($\sigma$ = 0.8, 1.0, and 1.4
m\,s$^{-1}$). Local trend fitting was then applied using the same time windows as LTF1,
but fitting the trends separately for each choice of random noise.

There are several features to note about this figure. The slope of the
power versus $N$ function is much steeper for the simulated data, even
with $\sigma$ as high as  1.4 m\,s$^{-1}$. LTF1 shows power that is
essentially flat except for the slight up-tick after the last data points are added.
Even then the FAP is $\approx$ 40\%. The power from LTF2 behaves  more erratically.
There is a sharp increase as the first data points are used, followed by just as sharp
a decline as more data are added. 
The ``high'' significance of the planet detection in LTF2 only occurs
after adding the last 100 data points which is 
inconsistent with what one expects for random noise,
This argues that the LTF1 choice of filtering may be a better approach to filtering out
the activity variations. 

We conclude that even though LTF2 shows modest power at the orbital frequency of the planet 
that this is not a significant detection and is consistent with the
non-detection of LTF1.

\subsection{The Influence of Noise}

\begin{figure}
\plotone{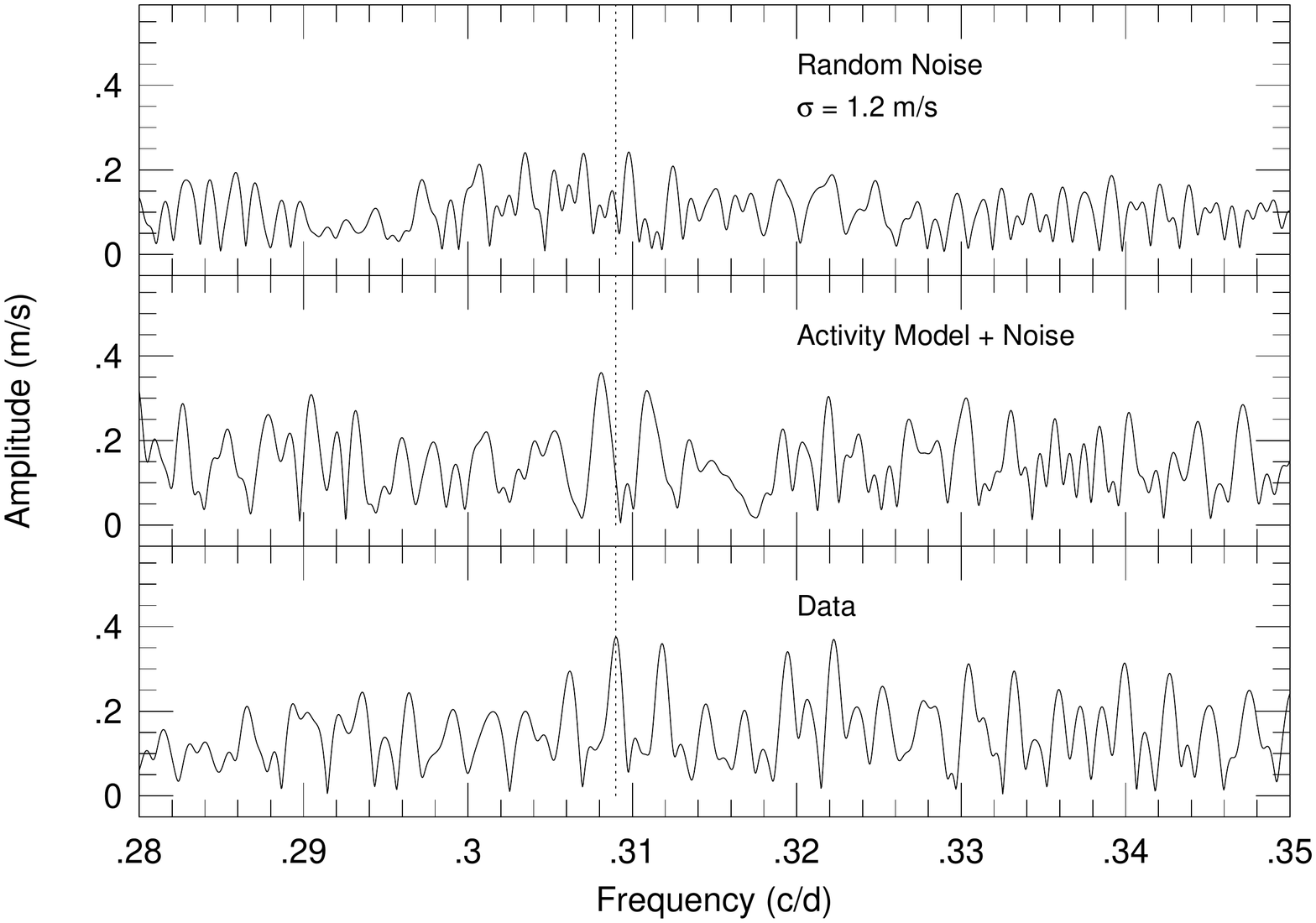}
\figcaption[]{(Top) The Fourier amplitude spectrum of random noise with $\sigma$ =
1.2 m\,s$^{-1}$ and with the same time sampling as the data. (middle) The amplitude
spectrum of the activity function with random noise
($\sigma$ = 1.2 m\,s$^{-1}$) added. (Bottom)
The Fourier amplitude spectrum of the actual unfiltered RV data.
The dashed vertical line marks the planet orbital frequency.
\label{noise}}
\end{figure}

The question naturally arises: 
``Why do some  approaches to removing the activity signal produce such discrepant results?''
The sampling coupled with the noise characteristics 
may gives us some insight into this. In this case it is best to compare the unfiltered data
in the Fourier domain.
The amplitude spectrum of random noise with $\sigma$ = 1.2 m\,s$^{-1}$, our 
worse case estimate of the noise level,
that is  sampled in the same way as the data shows several peaks 
near the planet frequency and with amplitudes comparable the $K$-amplitude of the planet
(top panel of Figure~\ref{noise}).  
Noise in the presence of the activity function 
shows an amplitude spectrum (middle panel of Figure~\ref{noise})
similar to that of the real data (lower panel of of Figure~\ref{noise}).
Spectral leakage from the activity signal into the frequency
range $\nu$ $\approx$ 0.3 -- 0.31\,d$^{-1}$ may  boost power in a noise
peak coincident with the planet orbital frequency. The details as to how this
noise peak is filtered may explain why the planet is present in some filtering
approaches, but not others. 

\begin{figure}
\plotone{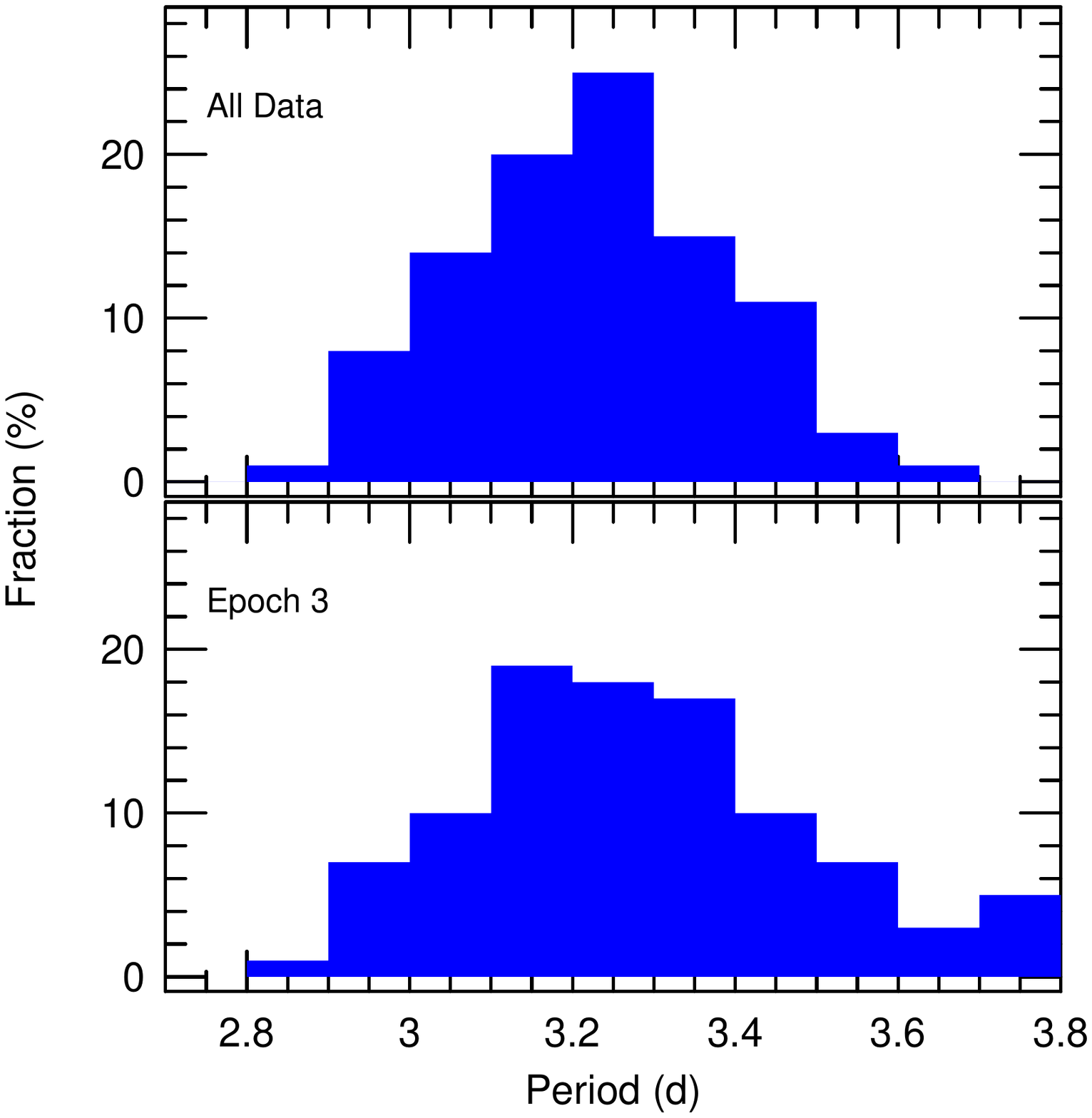}
\figcaption[]{(Top) The distribution of the dominant noise peaks in the period range
$P$ = 2.8 -- 3.8 \,d for the full data set. Random noise data sampled in the
same manner as the real data with $\sigma$ = 1.2 m\,s$^{-1}$ were used.
The mean amplitude of the noise peaks is $K$ = 0.24 $\pm$ 0.04 m\,s$^{-1}$.
(Bottom) The same as the top panel but using only the time sampling window of Epoch 3.
The mean amplitude of the noise peaks is $K$ = 0.33 $\pm$ 0.13 m\,s$^{-1}$.
\label{noisehist}}
\end{figure}

To explore further the influence of noise on the amplitude spectrum, synthetic
data consisting of only random noise (no activity signal)
with $\sigma$ = 1.2 m\,s$^{-1}$ were generated and
sampled in the same way as the real data. A total of 100 random data sets were created using 
different seed values for the random number generator. The top panel of 
Figure~\ref{noisehist} shows  the
distribution of the strongest peaks in the period range 2.85 days $<$ $P$ $<$ 3.8\,days
(0.26\,d$^{-1}$ $<$ $\nu$ $<$ 0.35\,d$^{-1}$). This has a peak 
at $P$ = 3.22 $\pm$ 0.01 \,days. The average amplitude of the peaks   is 
$K$ = 0.24 $\pm$ 0.04. In the unfiltered amplitude
spectrum the velocity amplitude at the planet orbital frequency is  0.38 m\,s$^{-1}$.  
The amplitude scales
approximately linear with $\sigma$, so for a noise level of 2 m\,s$^{-1}$ the noise
peaks would have an amplitude of $\approx$ 0.4 m\,s$^{-1}$.  

The epoch subsets of the data show similar noise characteristics. As an example we only
show the Epoch 3 data. The lower panel in
Figure~\ref{noisehist} show the noise peaks in the same frequency range for random data 
($\sigma$ = 1.2 m\,s$^{-1}$) sampled as the
Epoch 3 data. The distribution is similar, but in this case 
the average peak amplitude is slightly higher at $K$ = 0.33 $\pm$ 0.13 m\,s$^{-1}$.

As seen from Figure~\ref{noise} the activity signal may also boost the amplitude of the noise peak. The same simulation
was performed including the activity function  and random noise with $\sigma$ = 1.2 m\,s$^{-1}$.
In this case the noise peaks had a mean amplitude of $K$ = 0.39 $\pm$ 0.04,
essentially the same value as in the unfiltered amplitude spectrum.

\section{Discussion}

Alpha Cen B is a modestly active star which shows RV activity jitter 
with an amplitude $\approx$ 1.5 m\,s$^{-1}$ that is modulated
with the 38-days rotation period of the star. 
Our ability to extract reliably planetary signals with 
a much smaller amplitude depends on how well this activity is filtered out and whether
the filtering process introduces artifact frequencies.
Two methods were used to eliminate the activity variations from the RV
data of $\alpha$ Cen B:
traditional Fourier pre-whitening and local trend filtering (LTF). 
By using different approaches to filter the data, we hoped to obtain
a more accurate determination of the mass of $\alpha$ Cen Bb as well as to get
a better assessment of the statistical significance of the detected signal.

The significance of the 3.24\,days depends on how the activity variations are filtered.
Table 7 summarizes the FAP of the planet signal as determined
from the various filtering approaches applied in this paper. The values differ by factors
of 100, from a modestly significant detection to a non-detection. 
Interestingly, the method that gives lowest FAP, pre-whitening of the full data set,
is the one that has the poorest overall fit to the activity variations (see Figures~\ref{panel1} --
\ref{panel3}). Pre-whitening of the individual epochs produces a better fit to the underlying
activity (see Figure~\ref{epoch}) and consequently the FAP increases by almost a factor of 20.
Arguably the best fit to the activity signal, LTF1, produced no detection of the planet.
The planet $\alpha$ Cen Bb seems to be elusive -- for some filtering approaches it is there
while for others it is not.

The pre-whitening of the full RV data and LTF2 produced results that were most consistent with the D2012
result. In the case of the pre-whitened
we should assign little weight to the result as the fit to the 
overall activity variations is the poorest. There are some segments of the time series where
the fit is good, others where it is significantly poorer than for other methods.

The results for LTF2 cannot be so summarily set aside. Here almost the 
same procedure of LTF1 was followed, with the exception of the trend fits to 
four time chunks. Ostensibly the fit to the underlying activity
variations seems to as good as for LTF1. However, we have shown that the results depend 
sensitively on how the data of
Chunk8-9 are filtered. This chunk is problematic because there
are two large data gaps and the  RV data shows
complex variations. A Fourier analysis shows the presence of 3.3-d variations  which is 
dangerously close to the planet period. Different approaches to fitting the activity variations
in this chunk alone resulted in reduced power at the planet orbital frequency, and more importantly,
resulted  in higher power at a completely different frequency.
Because of the complex time variations in this chunk, the filtering approach of LTF1 is probably better.
Due to these difficulties the safest approach is to simply remove the data from Chunk8-9
from the analysis. Our simulations show that even without these
measurements the planet should have been detected with a much
higher statistical significance than was actually found for the data.

One could argue that since some methods for filtering out the activity variations do find
the planet signal that this qualifies as a confirmation of the presence
of $\alpha$ Cen Bb. However, a robust planet signal should be present at the
same level  regardless of how the data is analyzed. For  a weak  signal like $\alpha$ Cen Bb it is
difficult to judge which method one should trust (harmonic analysis, pre-whitening,
local trend fitting). We 
should emphasize that simulations have shown that if $\alpha$ Cen Bb were present according
to the orbit of D2012 {\it all} methods (pre-whitening of full data set,
pre-whitening of epoch data, and local trend filtering) should have detected the planet with high
significance (FAP $<$ 0.05 \%).

The analysis in Section 3.5 points to  noise as  a possible
explanation for the discrepant detections of $\alpha$ Cen Bb data. 
Simulations using random white noise with the ``worse case'' noise level of
1.2 m\,s$^{-1}$ that were sampled like the data can create peaks in the amplitude
spectrum near the period of the planet and with  a 
comparable amplitude ($K$ = 0.25 -- 0.4 m\,s$^{-1}$). The amplitude of these noise peaks
depends on two things: 1) the actual noise level of the RV data and its frequency spectrum,
and 2) the underlying activity variations and its frequency spectrum. 

The worse case noise level of $\sigma$ = 1.2 m\,s$^{-1}$ produces
a noise amplitude of $K$ = 0.24 m\,s$^{-1}$, but if the rms scatter is higher, the corresponding
amplitude increases. However, this estimate of the noise level in the $\alpha$ Cen B
RVs is derived {\it after} filtering out what we presumed were the activity variations. 
The unfiltered
RV data has an rms scatter of 2.1 m\,s$^{-1}$. If the true $\sigma$ is as high as 2 
m\,s$^{-1}$ the noise peak can have a value of 0.4 m\,s$^{-1}$, comparable to the unfiltered
$K$-amplitude of the  planet in the Fourier spectrum. 

The simulations also assumed Gaussian noise, but there are most likely systematic errors in the HARPS 
data and there is no guarantee that these also have a Gaussian distribution. 
Not only do we not know
the true rms scatter of this systematic noise, but most importantly we have
no knowledge of its  frequency characteristics.
Instead of being ``white'' the power spectrum
of the systematic noise maybe 
``red'' (i.e. overall slope rising to low frequencies), 
``blue'' (slope that rises to higher frequencies), or with strong peaks
(i.e. periodic signals in the data).
Given that 
observations are made on nightly, weekly, monthly, and yearly timescales 
it is reasonable to expect that periodic 
structure is present in the  Fourier amplitude spectrum
of the systematic noise. 
We have seen that white noise can produce peaks at the right frequency and amplitude
of the planet. We cannot be sure if systematic noise with a lower $\sigma$, but non-white
frequency structure  also boosts the amplitude of noise peaks in the Fourier spectrum
at the frequency of interest.
Perhaps LTF1 is a better way of filtering noise in the presence of the data window
which is why it produces no significant power at the planet orbital frequency.

The activity variations can also boost the amplitude of noise peaks. Using our simple
activity function and random noise with $\sigma$ = 1.2 m\,s$^{-1}$ we  were
able to produce
noise peaks with amplitudes consistent with the planet velocity $K$-amplitude.
The frequency spectrum of the activity signal for $\alpha$ Cen B is  certainly complex, one
that results from 
periodic and semi-periodic variations.
Intrinsic stellar variations  which 
can be stochastic (e.g. granulation, spot evolution, etc.) introduce ``noise'' with their own
frequency structure and this only complicates matters further. The activity signal, 
coupled with noise and the sampling window may 
also produce spurious peaks in the amplitude spectrum that may not be filtered out appropriately.
By lowering  the surrounding Fourier noise floor the filtering process may only make these spurious  
peaks look more significant than they really are.
For the detection of weak signals due to planets
it is essential to use different ways of filtering the data to ensure that we arrive at consistent
results. 

\section{Conclusions}

This investigation into the RV variations of $\alpha$ Cen B using  different
approaches to filtering out the activity signal was not able
to confirm the presence of the Earth-mass planet in a 3.24\,day orbit.
The detected  ``planet'' seemed to be highly sensitive to the details in 
how the activity variations are removed. 
Alpha Cen Bb should have been 
detected by all methods that were employed and at the same level of significance.
A possible explanation for the planet signal found by D2012
$\alpha$ Cen Bb is that it is a noise peak in the data whose statistical power 
has been boosted by 
a combination of the frequency
characteristics of the noise, the under-sampling of the activity signal, and the filtering process. 
This work cannot prove unequivocally that the RV signal attributed to $\alpha$ Cen Bb is
in fact noise. More 
analyses are needed, and most importantly more data taken should be taken with higher cadence so
as to sample adequately the activity variations. Only when the signal of $\alpha$ Cen Bb rises
with certainty above the noise level will we be certain of this planet.

In detecting the RV variations of low-mass exoplanets in the presence of activity noise
it is essential to confirm the detection using different filtering approaches.
We have shown that  standard Fourier pre-whitening can be a useful tool for finding
such signals, but the result should be verified using a different approach that 
is tailored to detecting the  frequency of interest. It should not be used indiscriminately.

Finally, a formal low false alarm probability is no guarantee that a periodic signal in RV
data is in fact significant, particularly when one has modified the data through
a filtering process. Simulated data should be used to ensure that the signal
was detected at the proper level and that its significance behaves in the expected manner
given the best estimate of the noise characteristics of the data. Quoted low values of the FAP
for activity filtered data should be treated with caution.

On a positive note, our analysis also shows that it is possible to extract the 
RV signal of
a short-period Earth-mass planet in the presence or activity noise given
exquisite quality data such as those taken with HARPS. However, high cadence observations
are required.
\vskip 0.1in

\centerline{Acknowledgments}

The author wishes to thank Bill Cochran, Mike Endl, and Guenther Wuchterl for 
useful comments on the manuscript. We also thank the anonymous referee for her/his careful
review and suggestions. This resulted in a much improved manuscript.

\clearpage

\begin{table}
\centering
\caption{Epochs of the HARPS data}
\label{epochtab}
\begin{tabular}{clllc}\hline
\hline
         & Dates            &  $\Delta$T & $N_{obs}$ & $\sigma$ \\
Epoch    & JD$-$2400000     & (days)     &           & (m\,s$^{-1}$) \\
\hline
1  & 54524.9--54648.6 &  124 & 42  & 1.15 \\
2  & 54878.8--55048.6 &  169 & 243 & 1.86 \\
3  & 54278.7--55359.5 &   91 & 120 & 2.42 \\
4  & 55611.8--55755.5 &  144 & 154 & 2.20 \\
\hline
\end{tabular}
\end{table}

\begin{table}
\centering
\caption{Pre-whitening results for the RV data}
\label{tlab}
\begin{tabular}{ccccc}\hline
N & Frequency   & Period  & $K$-amplitude  & Phase \\
  & (d$^{-1}$)  & (days)  & (m\,s$^{-1}$) & \\
\hline
\hline
f$_1$ & 0.02558 $\pm$  0.00002 & 39.09  $\pm$ 0.03    & 1.89 $\pm$ 0.08  & 0.00 $\pm$ 0.01 \\
f$_2$ & 0.00131 $\pm$  0.00005 & 763.36 $\pm$ 29.12   & 0.69 $\pm$ 0.08  & 0.79 $\pm$ 0.01 \\
f$_3$ & 0.08163 $\pm$  0.00004 & 12.25   $\pm$ 0.006 & 1.00 $\pm$ 0.08 & 0.49 $\pm$ 0.01 \\
f$_4$ & 0.10438 $\pm$  0.00005 &  9.58   $\pm$ 0.005 & 0.75 $\pm$ 0.08 & 0.13 $\pm$ 0.02 \\
f$_5$ & 0.00603 $\pm$  0.00004 & 165.83  $\pm$ 1.10 & 0.97 $\pm$ 0.08 & 0.78 $\pm$ 0.02 \\
f$_6$ & 0.06633 $\pm$  0.00005 & 15.79   $\pm$ 0.01 & 0.71 $\pm$ 0.08 & 0.57 $\pm$ 0.02 \\
f$_7$ & 0.03321 $\pm$  0.00005 & 101.11  $\pm$ 0.15 & 0.67 $\pm$ 0.09  & 0.62 $\pm$ 0.02 \\
f$_8$ & 0.07841 $\pm$  0.00005 & 12.75   $\pm$ 0.047 & 0.77 $\pm$ 0.08  & 0.05 $\pm$ 0.03 \\
f$_9$ & 0.30906 $\pm$  0.00009 & 3.2356  $\pm$ 0.0001 & 0.40 $\pm$ 0.08  & 0.33 $\pm$ 0.03\\
\hline
\end{tabular}
\end{table}
\clearpage

\begin{table}
\centering
\caption{Tests of the Fourier procedure on the full data set}
\label{pwtest}
\begin{tabular}{cccccc}\hline
\hline
$P_{in}$  &  $K_{in}$       &  $P_{out}$ & $K_{out}$  & FAP \\
(days)    &  (m\,s$^{-1}$)  &  (days)    &  (m\,s$^{-1}$) &   \\
\hline
3.350  &  0.50 &  3.349 $\pm$ 0.001 & 0.54 $\pm$ 0.08 & $<$5.0 $\times$ 10$^{-5}$   \\
3.300  &  0.50 &  3.299 $\pm$ 0.001 & 0.64 $\pm$ 0.09 & $<$5.0 $\times$ 10$^{-5}$  \\
3.236  &  0.50 &  3.236 $\pm$ 0.001 & 0.52 $\pm$ 0.08 & 1.5 $\times$ 10$^{-5}$   \\
3.200  &  0.50 &  3.203 $\pm$ 0.001 & 0.54 $\pm$ 0.10 & 2.0 $\times$ 10$^{-5}$   \\
3.150  &  0.50 &  3.151 $\pm$ 0.001 & 0.58 $\pm$ 0.10 & $<$5.0 $\times$ 10$^{-5}$  \\
\hline
\end{tabular}
\end{table} 
\clearpage

\begin{table}
\centering
\caption{Pre-whitening results for the Epoch data}
\label{epochpw}
\begin{tabular}{ccccc}\hline
\hline
ID & Frequency    & $K$-amplitude  & Phase & Comment \\ 
   & (d$^{-1}$)   & (m\,s$^{-1}$)  &       & \\
\hline
$f_{11}$ & 0.020  $\pm$ 0.001  &   0.89 $\pm$ 0.29 & 0.91 $\pm$ 0.04   & Epoch 1 \\
\hline
$f_{21}$ & 0.0043 $\pm$ 0.001 &   2.01 $\pm$ 0.15 & 0.27 $\pm$ 0.02   & Epoch 2   \\
$f_{22}$ & 0.0692 $\pm$ 0.001 &   0.76 $\pm$ 0.15 & 0.82 $\pm$ 0.03   & Epoch 2   \\
$f_{23}$ & 0.1347 $\pm$ 0.001 &   0.71 $\pm$ 0.15 & 0.07 $\pm$ 0.03   & Epoch 2   \\
\hline
$f_{31}$ & 0.0104  $\pm$ 0.0002 &   1.73 $\pm$ 0.16 & 0.33 $\pm$ 0.02   & Epoch 3   \\
$f_{32}$ & 0.0284 $\pm$  0.0003 &   2.52 $\pm$ 0.16 & 0.88 $\pm$ 0.02   & Epoch 3   \\
$f_{33}$ & 0.0644 $\pm$  0.0016 &   0.76 $\pm$ 0.16 & 0.38 $\pm$ 0.04   & Epoch 3   \\
$f_{34}$ & 0.1510 $\pm$  0.0016 &   0.69 $\pm$ 0.16 & 0.60 $\pm$ 0.04   & Epoch 3   \\
\hline
$f_{41}$ & 0.0267 $\pm$ 0.0005 &   1.59 $\pm$ 0.15 & 0.21 $\pm$ 0.02   & Epoch 4   \\
$f_{42}$ & 0.0818 $\pm$ 0.0005 &   1.63 $\pm$ 0.13 & 0.65 $\pm$ 0.02   & Epoch 4   \\
$f_{43}$ & 0.1028 $\pm$ 0.0014 &   1.63 $\pm$ 0.14 & 0.73 $\pm$ 0.04   & Epoch 4   \\
\hline
\end{tabular}
\end{table}
\clearpage

\begin{table}
\centering
\caption{Time chunks used for the Local Trend Filtering}
\label{ltf}
\begin{tabular}{ccccccc}\hline
\hline
Chunk \#   & Dates              & $\Delta$T  &  $N_{orb}$ &  $N_{data}$ & Fitting Function & $\sigma$ \\
           & (JD$-$2400000)     & (days)     &            &             &          & (m\,s$^{-1}$) \\
\hline
1 & 54524.91 -- 54530.83  & 5.93       &   1.83     &   7   & constant       &  0.53  \\
2 & 54548.82 -- 54557.76  & 8.94       &   2.16     &   10  & linear         &  0.57  \\
3 & 54562.78 -- 54571.72  & 8.90       &   2.75     &   9   & linear         &  1.19 \\
4 & 54610.72 -- 54617.62  & 6.88       &   2.13     &   5  & constant       &  1.13 \\
5 & 54638.69 -- 54648.64  & 9.95       &   2.99     &   11  & constant       &  1.06 \\
6 & 54878.80 -- 54884.89  & 6.09       &   1.86     &   10  & polynomial n=2 &  0.39 \\
7 & 54913.77 -- 54920.90  & 7.73       &   3.01     &   21  & polynomial n=2 &  0.55 \\
8 & 54933.70 -- 54938.82  & 5.12       &   1.58     &   18  & polynomial n=2 &  0.78 \\
9 & 54946.67 -- 54956.82  & 10.15      &   3.14     &   30  & polynomial n=3 &  0.78 \\
10 & 54988.53 -- 55002.68  & 14.15      &   4.34     &   29  & linear         &  0.84 \\
11 & 55036.55 -- 55048.58  & 12.04      &   3.72     &   18  & linear         &  1.10 \\
12 & 55278.74 -- 55301.88  & 23.14      &   7.15     &   62  & polynomial n=3 &  0.90 \\
13 & 55321.60 -- 55328.77  & 7.17       &   2.20     &   21  & polynomial n=2 &  0.84 \\
14 & 55334.59 -- 55342.76  & 8.17       &   2.52     &   19  & polynomial n=2 &  1.07 \\
15 & 55350.66 -- 55355.59  & 4.93       &   1.52     &   11  & polynomial n=2 &  0.95 \\
16 & 55611.79 -- 55616.80  & 5.00       &   1.23     &    9  & constant       &  0.80 \\
18  & 55619.77 -- 55648.84  & 29.07      &   8.98     &    44 & polynomial n=2 &  0.82 \\
19  & 55656.67 -- 55663.81  & 7.14       &   2.21     &    37 & polynomial n=2 &  1.04 \\
20  & 55672.65 -- 55692.70  & 9.95       &   3.08     &    36 & multi-sine  & 1.32 \\
21  & 55711.57 -- 55728.53  & 16.95      &   5.24     &    11 & linear	      &  1.58 \\
\hline
\end{tabular}
\end{table}
\clearpage

\begin{table}
\centering
\caption{Pre-whitening results for Chunk8-9}
\label{chunk89}
\begin{tabular}{cccc}
\hline\hline
Frequency    & Period   & $K$-amplitude  & Phase \\ 
(d$^{-1}$)& (days)   & (m\,s$^{-1}$) & \\
\hline
0.3020 $\pm$ 0.0034 & 3.31  $\pm$ 0.03 &  0.963 $\pm$ 0.15 & 0.77 $\pm$ 0.02 \\
0.1321 $\pm$ 0.0036 & 7.57  $\pm$ 0.21 &  0.902 $\pm$ 0.15 & 0.68 $\pm$ 0.02 \\
0.0271 $\pm$ 0.0070 & 36.90 $\pm$ 9.56 &  0.463 $\pm$ 0.15 & 0.64 $\pm$ 0.05 \\
\hline
\end{tabular}
\end{table}
\clearpage

\begin{table}
\centering
\caption{FAP values for the planet signal}
\label{faps}
\begin{tabular}{lc}\hline
\hline
Filter       &  FAP   \\
\hline
Full data Pre-whitening & 0.004  \\
Epoch Pre-whitening   & 0.07   \\
Local Trend fitting (LFT1)  &  0.40  \\
Local Trend fitting (LFT2)  & 0.005 \\
LFT2: various filters to  Chunk8-9 & 0.03 -- 0.3 \\
\hline
\end{tabular}
\end{table}
\clearpage

\clearpage


\clearpage

\clearpage

\clearpage

\clearpage

\clearpage

\clearpage

\clearpage
\clearpage
\clearpage

\clearpage

\clearpage

\clearpage


\clearpage

\clearpage

\clearpage

\clearpage

\end{document}